\documentclass[aps,pra,superscriptaddress,nofootinbib,showpacs,floatfix,twocolumn,notitlepage,longbibliography]{revtex4-2}

\usepackage{graphicx}
\usepackage{epstopdf}
\usepackage{amsmath}
\usepackage{amssymb}
\usepackage{mathrsfs}

\usepackage{hyperref}
\usepackage{bm}  
\usepackage{subfigure}

\hypersetup{
    colorlinks=true,       
    linkcolor=cyan,          
    citecolor=magenta,        
    filecolor=magenta,      
    urlcolor=cyan,           
    runcolor=cyan
}
\usepackage[pdftex]{color}

\newcommand{\beq}{\begin{equation}}
\newcommand{\eeq}{\end{equation}}
\newcommand{\beqnn}{\begin{equation*}}
\newcommand{\eeqnn}{\end{equation*}}
\newcommand{\bea}{\begin{eqnarray}}
\newcommand{\eea}{\end{eqnarray}}
\newcommand{\beann}{\begin{eqnarray*}}
\newcommand{\eeann}{\end{eqnarray*}}
\newcommand{\bes} {\begin{subequations}}
\newcommand{\ees} {\end{subequations}}

\newcommand{\ket}[1]{ \left| #1 \right \rangle}

\newcommand{\ketbra}[2]{|#1\rangle\langle #2|}

\newcommand{\ident}{\openone}
\newcommand{\Tr}{\mathrm{Tr}}

\newcommand{\ignore}[1]{}

\newcommand{\ev}[1]{\mathbb{E} \left[#1 \right]}

\usepackage{tikz}
\usetikzlibrary{quantikz2}
\usepackage{adjustbox}

\usepackage{mdframed}
\usepackage{tabularray}

\begin{document}
\title{Simulating the Open System Dynamics of Multiple Exchange-Only Qubits using Subspace Monte Carlo}

\author{Tameem Albash}
\email{tnalbas@sandia.gov}
\affiliation{Center for Computing Research, Sandia National Laboratories, Albuquerque NM, 87185 USA}

\author{N. Tobias Jacobson}
\email{ntjacob@sandia.gov}
\affiliation{Center for Computing Research, Sandia National Laboratories, Albuquerque NM, 87185 USA}
\begin{abstract}
We propose a Monte Carlo based method for simulating the open system dynamics of multiple exchange-only (EO) qubits. In the EO encoding, the total spin projection quantum number along the $z$-axis of the three constituent spins remains unchanged under exchange operations, in contrast to the open system (or multi-qubit miscalibration) setting where coherent and incoherent mixing of states with different {quantum numbers} occurs. In our approach, we choose to measure the {total spin component along the $z$-axis} of each EO qubit after every logical quantum operation, which decoheres coherent mixtures of states with different {spin projection quantum numbers}. 
Independent simulations thus give different trajectories of the system in the associated subspaces, so we refer to this method as the Subspace Monte Carlo method. With each EO qubit having a definite {spin projection quantum number}, the density matrix of $n$ qubits can be represented by a vector of dimension $3^{2n}$, instead of $8^{2n}$, with an additional vector of dimension $n$ to label the {quantum number} of each qubit. 
We show that this approximation of the dynamics remains faithful to the true dynamics when the simulated circuits twirl the noise, converting coherent errors to stochastic errors, which can be achieved using randomized compiling. 
We use this simulation approach to study how correlations in measurement outcomes of circuits with reset-if-leaked gadgets, such as a multi-round Bell state stabilization circuit that uses 6 EO qubits, are affected by the choice of CNOT implementations.
\end{abstract}
\maketitle

\section{Introduction}
Spin-full charge carriers confined to semiconductor quantum dots are a promising platform for scalable quantum information processors because of their compatibility with industrial semiconductor fabrication~\cite{Burkard2023,Steinacker2025}. The platform allows for different realizations of a qubit, from using a single spin (Loss-Divicenzo encoding~\cite{Loss1998,Kawakami2016,Yoneda2018,Zajac2018,Watson2018,Huang2018}), two spins (single-triplet encoding~\cite{Levy2002,Petta2005}), or more~\cite{Bacon2000,DiVincenzo2000,Kempe2001,Weinstein2005,Foulk2025,Shi2012,Taylor2013,Laird2010,Medford2013,Koh2012,Cao2016}, with varying requirements to enable universal control. 
The 3-spin exchange-only (EO) qubit~\cite{Bacon2000,DiVincenzo2000,Kempe2001,Laird2010,Eng2015,Russ2017} is an encoding that relies solely on exchange interactions between pairs of spins to enact universal quantum logic, with the added benefit of being unaffected by fluctuations in the global magnetic field. For semiconductor quantum dots, this exchange interaction can be turned on using baseband voltage control, making the EO qubit an attractive encoding for large-scale devices because of its simple control requirements. This qubit has been experimentally demonstrated and exhibits good fidelities~\cite{Andrews2019,Weinstein2023,Acuna2024,Madzik2025}.

The use of 3 spins to encode a single qubit poses a challenge for simulating EO qubits because the spin Hilbert space for a system of $n$ EO qubits grows as $8^n$, making multi-EO qubit simulations at the level of the spins with $n>2$ challenging. This can be further exacerbated if open system effects are taken into account, with the density matrix scaling as $8^{2n}$. This limits the kind of detailed noise modeling needed to inform our understanding of multi-EO qubit device performance in realistic noise environments, so approaches that enable larger scale simulations of EO qubits would be valuable.

In this work, we propose an approximation for simulating the dynamics of EO qubits in the presence of magnetic noise along the global field direction and voltage noise, where the vectorized density matrix scales as $3^{2n}$ with an additional vector label of size $n$. This reduced memory requirement is achieved by projecting each EO qubit (of the multi-qubit state) onto a subspace with a definite {spin projection quantum number along the $z$-axis}.
%
This projection is performed after each 1- or 2-qubit logical quantum operation, which remains a faithful simulation of the noisy dynamics of the constituent spins, so the projection forces the multi-qubit state before and after every operation to be expressible as a tensor product of single EO qubit states where each qubit has a definite subspace label. Under noiseless qubit operations with non-leakage input states, this projection does not affect the dynamics and only becomes relevant in the presence of noise and input leakage states. Because the outcome of this projection is non-deterministic and results in different trajectories for the dynamics, our simulations require averaging over different trajectories, which we implement using the Monte Carlo method~\cite{Hastings1970}. Under this averaging, coherent mixtures of the subspaces are decohered to classical mixtures.

To further improve the simulation efficiency, we use temporal coarse graining~\cite{Albash2025} to implement our 1- and 2-qubit gates, and we show simulation results for quantum circuits with up to 6 EO qubits. We use 2 EO qubit simulations to validate our method, which include 2-qubit simulations of repeated SWAP operations, 2-qubit randomized benchmarking~\cite{Magesan2011}, and 1-qubit randomized benchmarking with leakage detection after every Clifford operation. We find that our approximation remains faithful to the true dynamics when the circuits twirl the noise~\cite{Bennett1996a,Bennett1996b}, converting coherent errors to stochastic errors. This is naturally realized in randomized benchmarking circuits~\cite{Knill2008,Magesan2011} or more generally using randomized compiling~\cite{Wallman2016}.

As part of our benchmarking, we developed a randomized benchmarking protocol using reset-if-leaked (RiL) gadgets~\cite{Langrock2020} and single-qubit Clifford gates, where a RiL gadget is implemented after every Clifford gate as opposed to just at the end of the circuit. This protocol results in a similar decay curve for the survival probability as standard randomized benchmarking, while constantly resetting the qubit state to the ideal intermediate state in the circuit if leakage is detected. This has the advantage that no post-selection of the data needs to be performed. We believe this could be of independent interest as a protocol for characterizing RiL gadgets.

Beyond 2 qubits, we show 4-qubit simulations of 2-qubit randomized benchmarking with leakage detection and 6-qubit simulations of Bell state stabilization with leakage detection. These larger scale simulations enable us to study important dynamical properties of EO qubits, such as the probability of correlated leakage detection under different CNOT gate implementations.

The manuscript is organized as follows. In Sec.~\ref{sec:background}, we give an overview of the EO qubit encoding in order to establish the notation we use throughout. In Sec.~\ref{sec:NoiseH}, we define the magnetic and voltage noise we consider. In Sec.~\ref{sec:SubspaceMC}, we describe the Subspace MC method and our simulation algorithm, show simulations of 2 EO qubits that validate the method, and then show simulations with up to 6 EO qubits to highlight the approach to simulate multi-qubit systems. We conclude in Sec.~\ref{sec:conclusions}.

\section{Background for the Exchange-Only Qubit Encoding} \label{sec:background}
In the EO encoding of interest, a single qubit is encoded using three spins~\cite{Bacon2000,DiVincenzo2000,Kempe2001}. We choose the basis for the eight-dimensional Hilbert space of the three spins to be given by the simultaneous eigenstates of the commuting operators:
\bes \label{eqt:commutingoperators}
\begin{align}
\hat{S}_{123} &= \left( \sum_{i=1}^3 \vec{\hat{S}}^{(i)} \right) \cdot \left( \sum_{i=1}^3 \vec{\hat{S}}^{(i)} \right) \ , \\
\hat{S}_{12} & = \left( \sum_{i=1}^2 \vec{\hat{S}}^{(i)} \right) \cdot \left( \sum_{i=1}^2 \vec{\hat{S}}^{(i)} \right) \ , \\
\hat{S}_{z} &=  \sum_{i=1}^3 \hat{S}_z^{(i)} \ ,
\end{align}
\ees
where $\vec{\hat{S}}^{(i)} = \left( \hat{S}^{(i)}_x, \hat{S}^{(i)}_y, \hat{S}^{(i)}_z \right)$ is the vector of spin-1/2 operators along the $x,y,z$-axes. We denote the basis elements by $\ket{S, S_{12}, S_z}$ where $S, S_{12}, S_z$ are the quantum numbers associated with the operators in Eq.~\eqref{eqt:commutingoperators}, with $S \in \left\{1/2, 3/2 \right\}$, $S_{12} \in \left\{0, 1 \right\}$, such that the eight basis elements are given by:
\bes \label{eqt:EightBasisStates}
\begin{align}
\ket{0 _{-\frac{1}{2}}} &\equiv \ket{\frac{1}{2}, 0, -\frac{1}{2}} \ ,  \ket{0 _{\frac{1}{2}}} \equiv \ket{\frac{1}{2}, 0, \frac{1}{2}} \ , \label{eqt:State0}\\
\ket{1 _{-\frac{1}{2}}} &\equiv \ket{\frac{1}{2}, 1,  -\frac{1}{2}} \ ,  \ket{1 _{\frac{1}{2}}} \equiv \ket{\frac{1}{2}, 1, \frac{1}{2}} \ , \label{eqt:State1}\\
\ket{L _{-\frac{3}{2}}} &\equiv \ket{\frac{3}{2}, 1,  -\frac{3}{2}} \ ,  \ket{L _{-\frac{1}{2}}} \equiv \ket{\frac{3}{2}, 1,  -\frac{1}{2}} \ ,  \nonumber \\
\ket{L _{\frac{1}{2}}} &\equiv \ket{\frac{3}{2}, 1,  \frac{1}{2}} \ , \ket{L _{\frac{3}{2}}} \equiv \ket{\frac{3}{2}, 1,  \frac{3}{2}} \label{eqt:StateL} \ .
\end{align}
\ees
A decomposition of these states in terms of the spin basis $\left\{ \ket{\uparrow}, \ket{\downarrow} \right\}^{\otimes 3}$ is given in Appendix~\ref{app:BasisDecomposition}.

We assume the noiseless Hamiltonian of the three spins takes the form:
\beq \label{eqt:1qubitH}
\hat{H}_{\mathrm{NL}}(t) = g \mu_{\mathrm{B}} B_z   \sum_{i=1}^3 \hat{S}^{(i)}_z + \sum_{i=1}^2 J_{i,i+1}(t) \vec{\hat{S}}^{(i)} \cdot \vec{\hat{S}}^{(i+1)} \ ,
\eeq
where $B_z$ denotes a global magnetic field along the $z$-direction and where for simplicity we have assumed the spins are arranged in a line (a triangular arrangement is also possible~\cite{Acuna2024}). Any pulsing of the exchange interactions between the three spins preserves the quantum numbers $S$ and $S_z$ but allows for rotations between states with different $S_{12}$ value. For this reason, the $S_{12}$ index is used to label the computational basis index, and the $S_z$ value labels two different but equivalent computational subspaces. The four states in Eq.~\eqref{eqt:StateL} are states outside the computational subspace and are treated as leakage states. In the absence of leakage, a measurement of $S_{12}$ corresponds to a computational basis measurement.

%
The Hamiltonian (Eq.~\eqref{eqt:1qubitH}) restricted to either of the computational subspaces can be written as:
\begin{eqnarray} \label{eqt:logical1qubitH}
\left( \hat{H} \right)_{\mathrm{c.s.}} &=& - \frac{1}{4} (J_{12} + J_{23})  \hat{\ident} - \frac{1}{2} J_{12} \hat{\sigma}_z \nonumber \\
&& +\frac{1}{2} J_{23} \left( \sin(\phi) \hat{\sigma}_x  - \cos(\phi) \hat{\sigma}_z  \right) \ ,
\end{eqnarray}
where $\phi = 2 \pi/3$ and $\hat{\sigma}_{x,z}$ denote the $x,z$-Pauli operators on the computational basis states, e.g., $\hat{\sigma}_x \ket{0_{s_z}} = \ket{1_{s_z}}, \hat{\sigma}_z \ket{0_{s_z}} = \ket{0_{s_z}}$ for $s_z = \pm 1/2$. Thus, pulsing $J_{12}$ implements a clockwise rotation about the $z$-axis of the Bloch sphere, and pulsing $J_{23}$ implements a counter-clockwise rotation about the axis $\vec{n} = \left( \sqrt{3}/2, 0, 1/2 \right)$. These are then referred to as the Z and N exchange axes of the qubit. Because these axes are not parallel, pulsing $J_{12}$ and $J_{23}$ separately is enough to generate arbitrary rotations in the computational subspace. Another important consequence is that there is no dependence on the global magnetic field.

In practice, an EO qubit is initialized into the encoded $0$ state by preparing spins 1 and 2 in a singlet state ($\frac{1}{\sqrt{2}} \left( \ket{\uparrow_1 \downarrow_2} - \ket{\downarrow_1 \uparrow_2} \right)$. The state of the third spin is not set, so we can model the initial state as being a classical mixture of the two $S_z$ sectors:
\beq
\hat{\rho}_0 = p_{\frac{1}{2}}   \ketbra{0_{\frac{1}{2}}}{0_{\frac{1}{2}}} + p_{-\frac{1}{2}} \ketbra{0_{-\frac{1}{2}}}{0_{-\frac{1}{2}}}  \ ,
\eeq 
with $p_{\frac{1}{2}} \geq 0, p_{-\frac{1}{2}}\geq 0$ and $p_{\frac{1}{2}}  +p_{-\frac{1}{2}}  = 1$. Since we can interpret this density of state as an ensemble of pure states, in what follows, we model the initial state as a pure state that is randomly picked to be $\ket{0_{\frac{1}{2}}}$ with probability $p_{\frac{1}{2}}$ or $\ket{0_{-\frac{1}{2}}}$ with probability $p_{-\frac{1}{2}}$. 

The exchange interaction alone will not result in population transfer between the two $S_z = \pm 1/2$ sectors. For this reason, starting from an initial state of $\ket{0_{-\frac{1}{2}}}$ or $\ket{0_{\frac{1}{2}}}$, certain spin states cannot be reached and cannot represent logical states. To illustrate this point, we consider the following three-spin pure state
\beq
\ket{\psi} = \frac{1}{\sqrt{2}} \left( \ket{{0_{-\frac{1}{2}}}} + \ket{{1_{\frac{1}{2}}}} \right) \ .
\eeq
This state does \emph{not} correspond to an encoded state $\ket{+} = \frac{1}{\sqrt{2}} \left( \ket{0} + \ket{1} \right)$ and cannot be prepared using the Hamiltonian in Eq.~\eqref{eqt:logical1qubitH} starting from either of $\ket{{0_{-\frac{1}{2}}}}$ or $\ket{{0_{\frac{1}{2}}}}$. For example, if we were to apply a  Hadamard operation $\hat{U}_H$ to the state $\ket{\psi}$ (the implementation of encoded Clifford operations can be found in Ref.~\cite{Andrews2019}), we would get:
\beq
\hat{U}_{H} \ket{\psi} = \frac{1}{2} \left( \ket{0_{-\frac{1}{2}}}  +  \ket{1_{-\frac{1}{2}}} + \ket{0_{\frac{1}{2}}}- \ket{1_{\frac{1}{2}}} \right) \ .
\eeq
When performing a measurement of $S_{12}$, we would find equal probability of 0 and 1, which does not correspond to an encoded $\ket{0}$.

For two qubits or more, we write the basis as the tensor product of the single qubit bases, but there is a choice of ``orientation'' that we need to make depending on which spins are used for the $S_{12}$ quantum number, or equivalently which exchange interaction we associate with the $(-z,n)$ rotations for each qubit. For example, for 2 qubits, if we label the spins 1,2,3 for the first qubit and 4,5,6 for the second qubit and if we associate the $z$ rotations with the $J_{12}$ and $J_{56}$ exchange interactions, then the basis we choose is:
\beq
\left\{ \left| S^{(1)}, S_{12}^{(1)}, S_z^{(1)} \right \rangle \otimes  \left| S^{(2)}, S_{56}^{(2)}, S_z^{(2)} \right \rangle \right\} \ .
\eeq
We call this the ZNNZ orientation. Similarly, if we associate the $z$ rotations with the $J_{23}$ and $J_{45}$ exchange interactions, then the basis we choose is:
\beq
\left\{ \left| S^{(1)}, S_{23}^{(1)}, S_z^{(1)} \right \rangle \otimes  \left| S^{(2)}, S_{45}^{(2)}, S_z^{(2)} \right \rangle \right\} \ .
\eeq
We call this the NZZN orientation. Another commonly used orientation is the ZNZN orientation; see for example Ref.~\cite{Chadwick2025}.
\section{Noise Hamiltonian} \label{sec:NoiseH}
%
The total Hamiltonian $\hat{H}(t)$ is taken as a sum of the noiseless Hamiltonian (of the form in Eq.~\eqref{eqt:1qubitH}) and a noise Hamiltonian that includes magnetic and charge noise. We discuss the form of these two noise sources next.
%
\subsection{Magnetic noise} 
%
We assume the magnetic noise can be captured with the Hamiltonian:
\beq \label{eqt:MagneticNoise}
\hat{H}_{\mathrm{M}}(t) =  \sum_i \delta h_i(t) \hat{S}^{(i)}_z \ ,
\eeq
where each $\delta h_i$ is taken to be a sum of independent OU processes.
This choice of having only magnetic noise along the $z$-direction (the same direction as the global magnetic field in Eq.~\eqref{eqt:1qubitH}) has important consequences for us. For a single qubit, $\hat{S}^{(i)}_z$ only commutes with $\sum_{i} \hat{S}^{(i)}_z$ from the list in Eq.~\eqref{eqt:commutingoperators}, so in the presence of such magnetic noise, we can treat the dynamics in the $S_z = -3/2, -1/2, 1/2, 3/2$ subspaces separately. Similarly, for two qubits, we can treat the dynamics in the combined $S_z = S_z^{(1)} + S_z^{(2)}  = -3, -2, -1, 0, 1, 2, 3$ subspaces separately. 
{This assumption of magnetic noise only along $\hat{S}_{z}$ is justified in the regime of sufficiently large applied magnetic field~\cite{Ladd2012} based on a simple perturbation theory argument. The necessary field for this to be a good approximation may be small, in practice. For example, if the transversal magnetic noise strength $\delta B_{\perp}$ due to, e.g., nuclear Overhauser field fluctuations were to have a scale of a few kHz \cite{Veldhorst2014} ($\mathcal{O}(0.1 \ \mathrm{\mu T})$ based on the electron gyromagnetic ratio of $\sim \! 28 \ \mathrm{GHz/T}$), an applied field of order only $\mathcal{O}(100 \ \mathrm{\mu T})$ would be required such that the Overhauser field perturbation due to $\hat{S}_{x,y}$ components is only $\mathcal{O}(10^{-3})$ the magnitude of the $\hat{S}_{z}$ magnetic noise components.}
%
\subsection{Voltage noise} 
%
We model the exchange interaction as a function of applied voltage $V(t)$ with the Hamiltonian:
\begin{eqnarray}
\hat{H}_{\mathrm{EX}}(t) &=& J_0 e^{\frac{V_{ij}(t) + \delta V_{ij}(t)}{\mathcal{I}}} \vec{\hat{S}}^{(i)} \cdot \vec{\hat{S}}^{(j)} \nonumber \\
&& \hspace{-1.5cm} = \left( J_{ij}(t) + J_{ij}(t) \left( e^{\frac{ \delta V_{ij}(t)}{\mathcal{I}}} - 1 \right) \right)   \vec{\hat{S}}^{(i)} \cdot \vec{\hat{S}}^{(j)} \ ,
\end{eqnarray}
where $J_{ij}(t) = J_0 e^{\frac{V_{ij}(t)}{\mathcal{I}}}$, $\mathcal{I}$ is the insensitivity, and $\delta V_{ij}(t)$ is taken to be a sum of independent OU processes representing the effect of voltage noise. The treatment of the stochastic process $\delta J_{ij}(t) = J_{ij}(t) \left( e^{\frac{ \delta V_{ij}(t)}{\mathcal{I}}} - 1 \right) $, which is \emph{not} a sum of OU processes, is discussed in Appendix~\ref{app:ChargeNoise}. For simplicity, we assume that $J_{ij}(t) = 0$ when $V_{ij}(t) = 0$ so that we can ignore the effect of residual exchange when the voltage is off. This has the important consequence of allowing us to restrict ourselves to purely single-qubit and two-qubit operations.

Because the voltage noise only involves operators of the form $\vec{\hat{S}}^{(i)} \cdot \vec{\hat{S}}^{(j)}$, it preserves {the spin projection quantum number along the $z$-axis} $S_z$ for single qubit operations and the sum of the quantum numbers $S_z = S_z^{(1)} + S_z^{(2)}$ for two qubit operations.

\section{Subspace Monte Carlo} \label{sec:SubspaceMC}
%
\subsection{Approximation} \label{sec:SubspaceMCapprox}
Based on the assumptions established in Sec.~\ref{sec:NoiseH}, we can restrict our attention to logical quantum operations acting on at most 2 qubits. To describe our approach to model reduction, we begin by assuming that each qubit has a definite {spin projection quantum number along the $z$-axis}. This means that the reduced density matrix for the $i$-th EO qubit of the system can be expressed as:
\beq \label{eqt:1qubitszform}
\hat{\rho}^{(i)}_{s_z^{(i)}} = \sum_{s,s',s_{12},s_{12}'} a_{s,s',s_{12},s_{12}'}^{(i)}  \left| s , s_{12}, s_{z}^{(i)} \right \rangle \left \langle s',s_{12}',s_{z}^{(i)}  \right | \ ,
\eeq
for a fixed $s_{z}^{(i)}$. Because the value of $s_{z}^{(i)}$ picks out a unique set of computational and leakage states (see Eq.~\eqref{eqt:EightBasisStates}), we can compress the description of the single EO qubit state as:
\beq  \label{eqt:1qubitszform2}
\hat{\rho}^{(i)}_{s_z^{(i)}} = \sum_{x,y\in \left\{0,1,L\right\}} a_{x,y}^{(i)} \ketbra{x_{s_{z}^{(i)}}}{y_{s_{z}^{(i)}}} \ .
\eeq
We can therefore think of our 3-spin qubit as being effectively described by a qutrit (3-level system) with an additional {spin projection quantum number} index.

For a two-qubit operation acting on qubits $i$ and $j$, the dynamics preserves the sum of the {spin projection quantum numbers along the $z$-axis}, $s_{z}^{(i)} + s_{z}^{(j)}$, and we can expect the reduced density matrix for each qubit to no longer be in the form of Eq.~\eqref{eqt:1qubitszform} or Eq.~\eqref{eqt:1qubitszform2} for genuine two-qubit operations (for a tensor product of single-qubit operations, it would remain in this form). For the well-established implementations of two-qubit logical gates~\cite{Fong2011,Weinstein2023} and for input states in the computational subspace, a superposition or mixture of states with different spin projection quantum numbers arises solely from noise or imperfect implementation of the gates, so we can expect the degree of mixing to be small for a well-calibrated system with low noise. 

When the input state includes leakage states, the output can result in a superposition of states with different {spin projection quantum numbers}. This happens for non-leakage-controlled logical operations and for the RiL gadget~\cite{Langrock2020}. For example, if we consider the (not leakage controlled) CNOT gate of Ref.~\cite{Fong2011}, given an input state with a target qubit in a leaked state and the control qubit in the $\ket{0}$ state, $ \ket{L_{ \frac{1}{2}}} \ket{ 0_{\frac{1}{2}}}$ (NZ-ZN orientation), then the output state would be:
%
\begin{eqnarray} \label{eqt:LeakagePropagaion}
\hat{U}_{\mathrm{FWCX}} \ket{L_{\frac{1}{2}}} \ket{0_{\frac{1}{2}}} &=& \nonumber \\
&& \hspace{-2cm} -0.69  \ket{L_{\frac{1}{2}}} \ket{0_{\frac{1}{2}}} +  0.11  \ket{L_{\frac{1}{2}}} \ket{1_{\frac{1}{2}}} \nonumber \\
&& \hspace{-2cm}  +0.37 \ket{L_{\frac{3}{2}}} \ket{1_{ -\frac{1}{2}}} + 0.53 \ket{L_{ -\frac{1}{2}}} \ket{L_{\frac{3}{2}}} \nonumber \\
&&\hspace{-2cm}  -0.15  \ket{L_{\frac{1}{2}}} \ket{L_{\frac{1}{2}}} -0.27  \ket{L_{\frac{3}{2}}} \ket{L_{ -\frac{1}{2}}} \ , 
\end{eqnarray}
where the output state is a superposition of states with different {spin projection quantum numbers} for each qubit.

\begin{figure}
\includegraphics[width=1.5in]{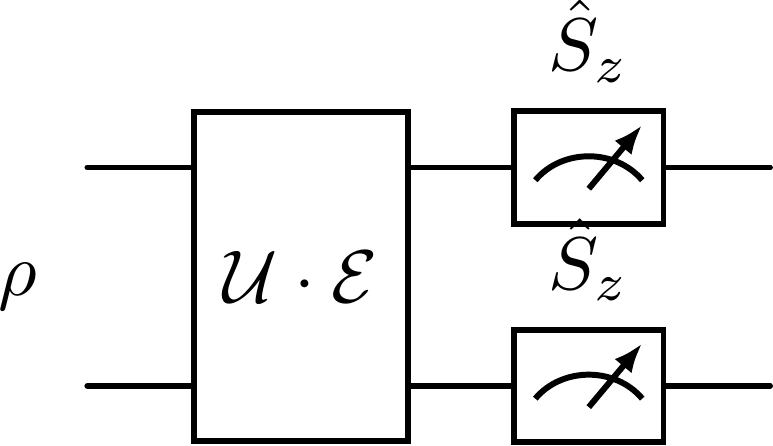} 
\caption{\textbf{Illustration of the Subspace MC approximation on two EO qubits.}  After the application of the logical quantum operation $\mathcal{U} \cdot \mathcal{E}$, a measurement of the $\hat{S}_z$ operator on each qubit is performed. This projects each qubit onto a subspace with a fixed $s_z$ value.} \label{fig:GaugeProjection}
\end{figure}

Because having a definite {spin projection quantum number} for each qubit allows us to compress the representation of the state (Eq.~\eqref{eqt:1qubitszform2}), our approach for model order reduction is to project each qubit onto a definite {spin projection quantum number} after each logical quantum operation. We can think of this as performing a measurement of the {spin component along the $z$-axis} of each qubit, as depicted in Fig.~\ref{fig:GaugeProjection}. The measurement outcome, and hence the projection, is probabilistic. Therefore, we expect that even with identical noise realizations, independent simulations can provide different trajectories for the state depending on which projection occurs. By averaging over these different simulations, we get a classical mixture of states with different spin projection quantum numbers and not a coherent superposition, which is the (uncontrolled) approximation being made. This approximation then allows us to express an $n$-EO qubit density matrix in a more compact form:
\begin{eqnarray} \label{eqt:SubspaceMCState}
\hat{\rho}_{s_{z}^{(1)}, \dots, s_{z}^{(n)}} &=& \sum_{x,y\in \left\{0,1,L\right\}^{\otimes n}} a_{x,y} \times \nonumber \\
&& \hspace{-2.0cm} \ketbra{(x_1)_{s_{z}^{(1)}} \dots (x_n)_{s_{z}^{(n)}}}{(y_1)_{s_{z}^{(1)}} \dots (y_n)_{s_{z}^{(n)}}} \ , 
\end{eqnarray}
which scales as $3^{2n} + n$, since we have a system of $n$ qutrits, with each qutrit having an additional $s_z^{(i)}$ label that we can update after each projection. This is to be contrasted with the scaling of $8^{2n}$ if we were to express the density matrix of the $n$-EO qubits in the spin basis.
%
\subsection{Algorithm}

In order to implement our simulations, we use the temporal coarse graining technique from Ref.~\cite{Albash2025} for each logical gate. In this approach, we replace the dynamics associated with the unitary over the entire duration of a logical gate 
\begin{eqnarray}
\hat{U}(t_1,t_0) &=& \mathrm{T} \exp \left[ -i \int_{t_0}^{t_1} dt  \left( g \mu_{\mathrm{B}} B_z  \sum_{i=1} \hat{S}^{(i)}_z \right. \right. \nonumber \\
&& \left. \left. + \hat{H}_{\mathrm{M}}(t) + \hat{H}_{\mathrm{EX}}(t) \right) \right] \ , 
\end{eqnarray}
by the quantum operation $\mathcal{U} \cdot \mathcal{E}$, where $\mathcal{U}$ is the unitary operation in the absence of noise $(\delta h_i = 0, \delta V_{ij}(t)  = 0)$ and $\mathcal{E}$ includes contributions from averaging over the OU bridge processes resulting in decoherence as well as unitary dynamics associated with the boundary values of the OU processes. We refer the reader to Ref.~\cite{Albash2025} for further details on this approach as well as a brief description in Appendix~\ref{app:TemporalCoarseGraining}. While this technique is general, in our simulations we use only 1- and 2-qubit logical gates, so the operation $\mathcal{U} \cdot \mathcal{E}$ acts on at most 6 spins.
 
If our input density matrix is of the form in Eq.~\eqref{eqt:SubspaceMCState}, we know that the initial input state to the operation $\mathcal{U} \cdot \mathcal{E}$ has a well-defined $s_z = s_z^{(1)} + s_z^{(2)}$ value, and the dynamics of $\mathcal{U} \cdot \mathcal{E}$ will preserve that value. This can be used to implement the operation in a subspace of smaller dimension. For example, if the 2-qubit input state has $s_z= 0$, then the operation acts on a subspace of dimension 20, which is smaller than using the 64 dimensional Hilbert space of 6 spins. 

After the application of $\mathcal{U} \cdot \mathcal{E}$, the output state may no longer be of the form of Eq.~\eqref{eqt:SubspaceMCState}, so we perform the projection described in the previous subsection and depicted in Fig.~\ref{fig:GaugeProjection}. The projected state is then in the form of Eq.~\eqref{eqt:SubspaceMCState}. Therefore, if the initial state of our quantum circuit is in the compressed form of Eq.~\eqref{eqt:SubspaceMCState}, then it will remain in this form throughout the simulation. Because the outcome of the projection is probabilistic, for a given magnetic and voltage noise realization, we perform independent simulations such that other projection outcomes are explored, and we average over these simulations. This gives our approach a Monte Carlo (MC) aspect beyond the averaging over independent noise realizations, which is why we refer to the simulation method as Subspace MC.

\subsection{Benchmarking}
%
In what follows, we use the implementation of Clifford gates from Ref.~\cite{Andrews2019}, a SWAP gate implemented with 15 exchange pulses, and two different implementations of CNOT gates. One is a leakage-controlled CNOT gate~\cite{Weinstein2023} implemented with 47 exchange pulses (LCCX), and the other is non-leakage controlled~\cite{Fong2011} implemented with 28 exchange pulses (FWCX). In the former case, a leaked state on either the control qubit or the target qubit does not propagate to the other, whereas in the former a leaked state on the target qubit  results in a superposition of computational and leaked states on the control qubit, as shown in Eq.~\eqref{eqt:LeakagePropagaion}. While there can be some variation in the number of exchange pulses depending on the orientation of the qubits (ZNNZ versus NZZN), here we have chosen to use the same number of pulses for both orientations for consistency. As some of our next examples highlight, there can be a tradeoff between preventing leakage spreading using the LCCX gate and reducing the circuit duration and hence the impact of magnetic and charge noise using the FWCX gate.

For systems of one or two qubits, we are able to compare the Subspace MC method against simulations where the subspace projection is not performed. We refer to the latter simulations as `Exact', and we use this comparison to benchmark the validity of the Subspace MC method. Here we ignore any error associated with the temporal coarse graining method. In the simulation results we present next, we use the noise parameters parameters labeled NM1 in Tables~\ref{table:magneticT2} and \ref{table:voltageT2} for the magnetic and voltage noise respectively. Furthermore, we define our computational basis measurement to have two POVM elements: $E_0 = \sum_{s_z=-1/2}^{1/2} \ketbra{0_{s_z}}{0_{s_z}}$ corresponding to outcome 0, and $E_1 = \sum_{s_z=-1/2}^{1/2} \ketbra{1_{s_z}}{1_{s_z}} + \sum_{s'_z=-3/2}^{3/2} \ketbra{L_{s'_z}}{L_{s'_z}}$ corresponding to outcome 1. We assume that the measurement is ideal in all other ways. For the initial state, all qubit states are initialized in the computational space with random $s_z \in \left\{ -1/2, 1/2\right\}$ values.

\subsubsection{2-qubit repeated SWAPs} \label{sec:Swap}
\begin{figure*}[t] 
   \centering
   \hspace{0.75in}
   \subfigure[]{\includegraphics[width=1.in]{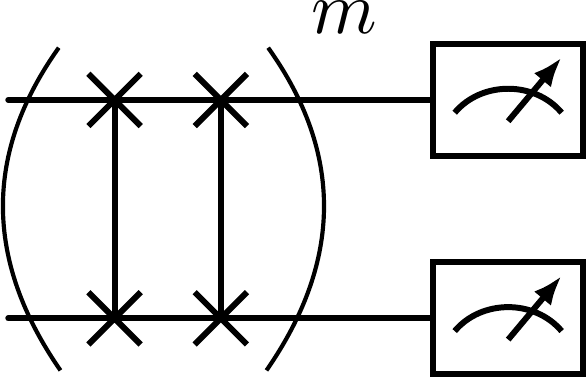} \label{fig:RepeatedSwapsCircuit}} 
   \hspace{1.25in}
   \subfigure[]{\includegraphics[width=2.25in]{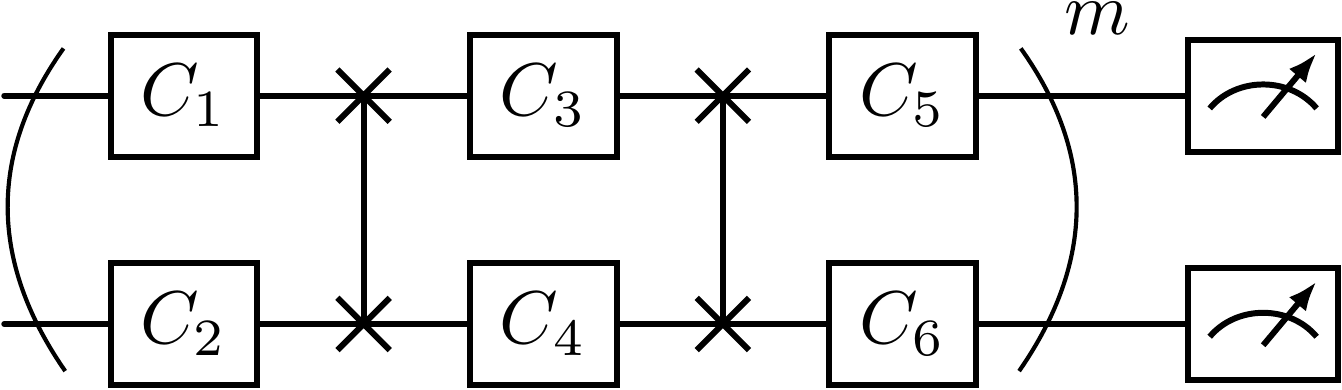} \label{fig:RepeatedRCSwapsCircuit}}
   \subfigure[]{\includegraphics[width=3in]{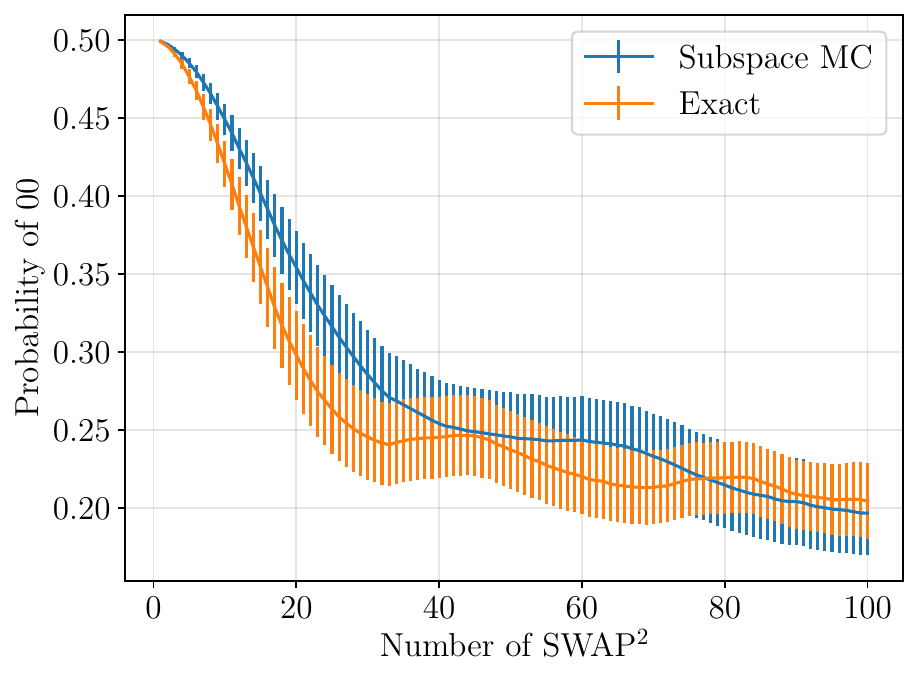} \label{fig:RepeatedSwapsResults}}
   \subfigure[]{\includegraphics[width=3in]{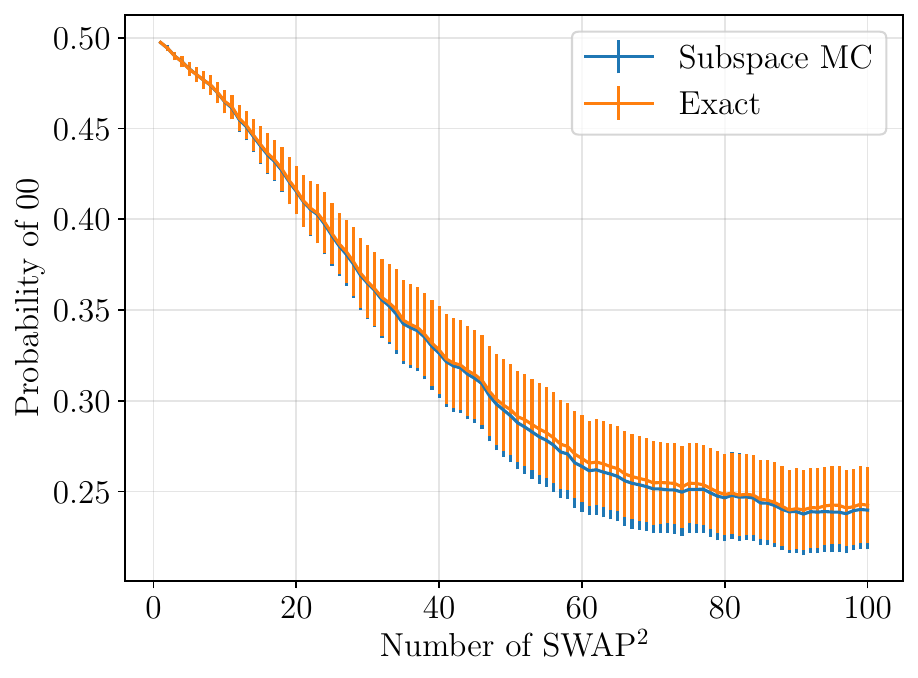} \label{fig:RepeatedRCSwapsResults}}
   \caption{\textbf{Repeated SWAP pairs.} (a) Illustration of the quantum circuit implementing $m$ repeated applications of a pair of SWAPs. The circuit is terminated by a computational basis measurement of each qubit. (b) Same as in (a), except we include randomized compiling for the pair of SWAP gates. (c) and (d) Probability of measuring the outcome 0 for both qubits  when simulating the circuit in (a) and (b) respectively when acting on the initial Bell state $\ket{\Phi} = \frac{1}{\sqrt{2}} \left( \ket{00} + \ket{11} \right)$. Simulation results are averaged over 100 noise realizations using noise model NM1, with the error bars being the $2\sigma$ confidence intervals generated by performing a bootstrap over the 100 noise realizations. For the Subspace MC simulations, 50 MC trials are performed per noise realization.}
\end{figure*}

%
As indicated in Sec.~\ref{sec:SubspaceMCapprox}, we expect the Subspace MC approach to fail in cases where coherent errors amplify under repeated applications of gates. We illustrate this by repeatedly applying pairs of SWAP operations, which in the absence of noise should implement the identity operation. We take the initial state to be the Bell state $\ket{\Phi} = \frac{1}{\sqrt{2}} \left( \ket{00} + \ket{11} \right)$ with random {spin projection quantum numbers} chosen for each qubit. The circuit is shown in Fig.~\ref{fig:RepeatedSwapsCircuit}. In Fig.~\ref{fig:RepeatedSwapsResults}, we show a comparison between the exact evolution and that of Subspace MC for measuring outcome 00 with increasing number of pairs of SWAPs. While both show a decline in the probability, the Subspace MC result departs from the exact simulation very early on.

As noted earlier, the disagreement between the exact and Subspace MC simulations arises because the build-up of noise-induced coherent errors from multiple applications of the SWAP are converted into incoherent errors by the subspace projection. To restore the agreement, we implement randomized compiling~\cite{Wallman2016} in our circuit. This is illustrated in Fig.~\ref{fig:RepeatedRCSwapsCircuit}, where we choose random single-qubit Clifford generators $C_1, \dots, C_6$ at every pair of SWAP operations, such that the grouping is logically equivalent to an identity. We note that we could combine neighboring single-qubit Clifford operations between applications into a single Clifford operation, but we do not do this here and it does not affect our conclusions. In Fig.~\ref{fig:RepeatedRCSwapsResults}, we show how the simulation results of the exact and Subspace MC simulations for the randomized compiled circuits are in agreement within statistical confidence. We get similar agreement if we use Pauli gates instead of the Clifford gates, as we show in Appendix~\ref{app:MoreResults}.

\begin{figure*}[t] 
   \centering
   \hspace{0.75in}
   \subfigure[]{\includegraphics[width=1.in]{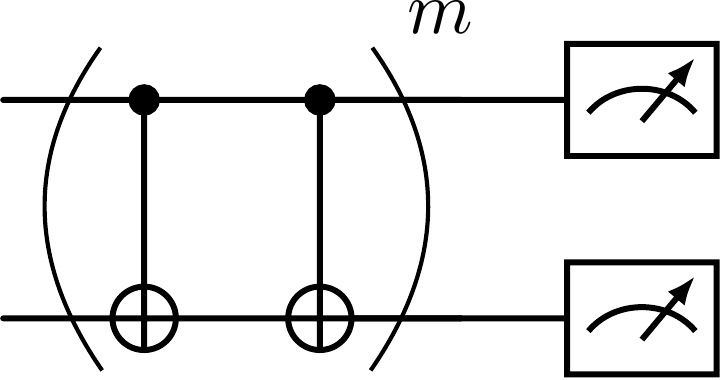} \label{fig:RepeatedCXCircuit}} 
   \hspace{1.25in}
   \subfigure[]{\includegraphics[width=2.25in]{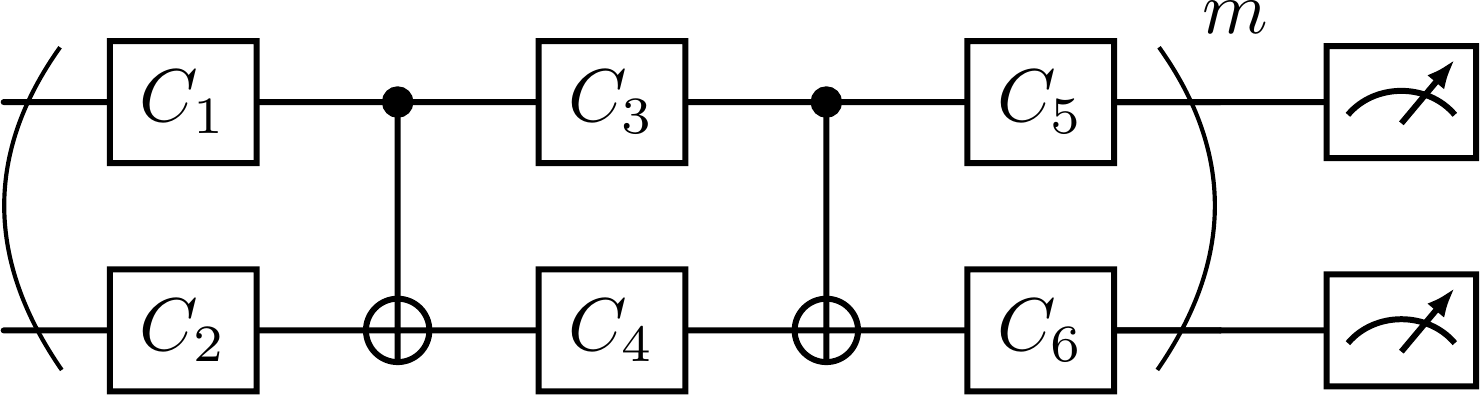} \label{fig:RepeatedRCCXCircuit}}
   \subfigure[]{\includegraphics[width=3in]{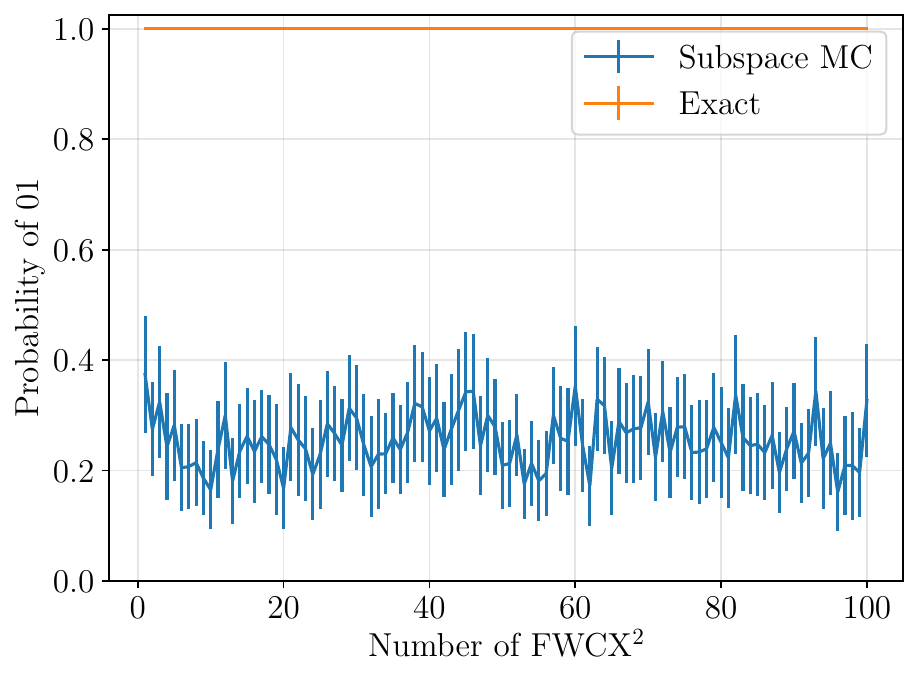} \label{fig:RepeatedFWCXResults}}
   \subfigure[]{\includegraphics[width=3in]{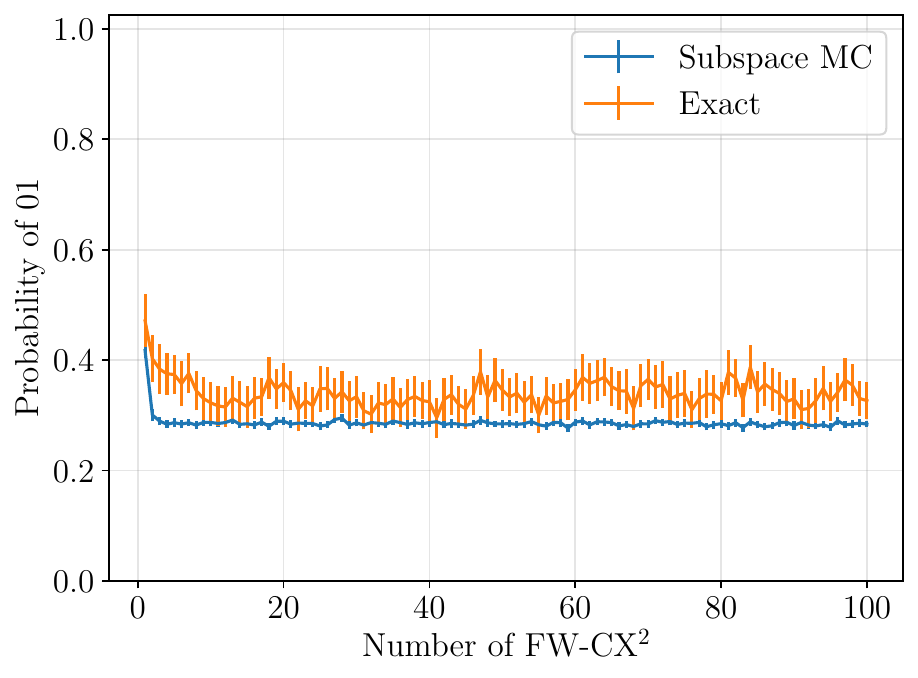} \label{fig:RepeatedRCFWCXResults}}
   \caption{\textbf{Repeated FWCX pairs.} (a) Illustration of the quantum circuit implementing $m$ repeated applications of a pair of FWCX gates. The circuit is terminated by a computational basis measurement of each qubit. (b) Same as in (a), except we include randomized compiling for the pair of FWCX gates. (c) and (d) Probability of measuring the outcome 0 and 1 for each qubit respectively when simulating the circuit in (a) and (b) when acting on the initial state $ \ket{0_{1/2} L_{1/2}}$. Simulations involve no noise. For the Subspace MC simulations in (c), the error bars are the $2\sigma$ confidence intervals generated by performing a bootstrap over 50 independent MC realizations. For the simulations in (d), the error bars are the $2\sigma$ confidence intervals generated by performing a bootstrap over 100 independent randomized compiling realizations. For the Subspace MC simulations, 50 MC trials are performed per randomized compiling.}
\end{figure*}

\begin{figure}[htbp] 
   \centering
   {\includegraphics[width=3.25in]{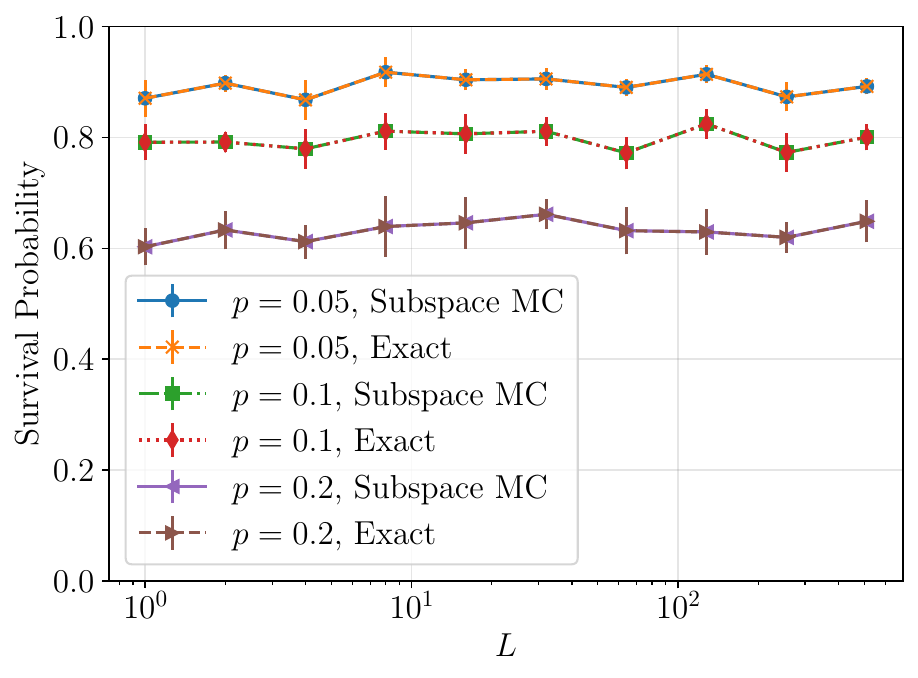} }
   \caption{\textbf{Two-qubit RB with only state preparation errors.} Average survival probability as a function of circuit depth $L$ for 2-qubit randomized benchmarking using the FWCX gate. At every depth, 50 random circuits are averaged over. 10 independent noise realizations for the state preparation are generated, and the two simulation methods presented use the same random seed for the state preparation. For the subspace MC simulations, 5 Monte Carlo trials are performed per circuit. Error bars are $2\sigma$ confidence intervals generated by performing a bootstrap over the 10 noise realizations.}
   \label{fig:2QRB_stateprep}
\end{figure}

\begin{figure}[htbp] 
   \centering
   {\includegraphics[width=3.25in]{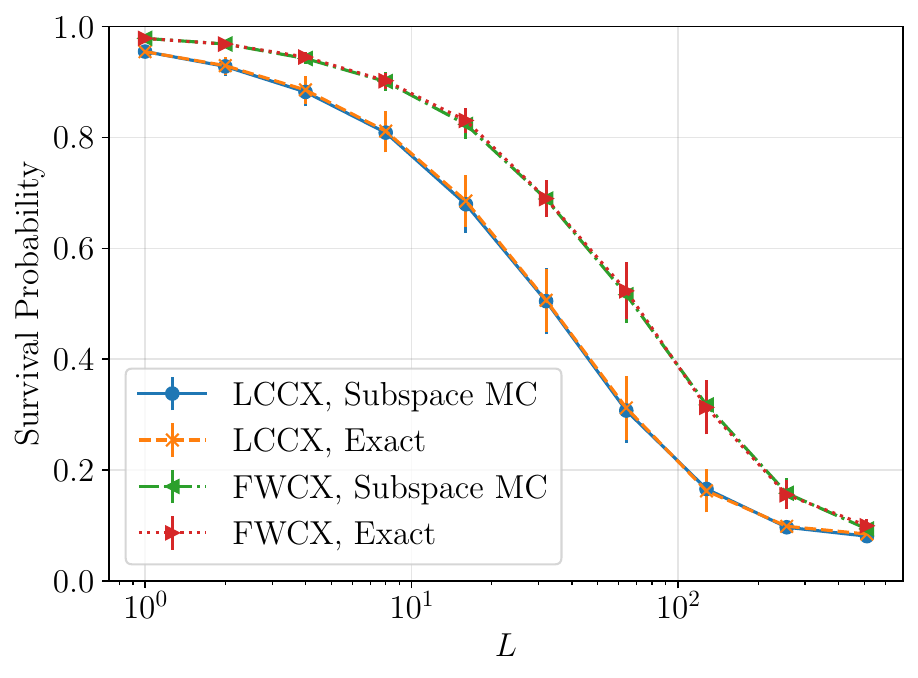} }
   \caption{\textbf{Two-qubit RB.} Average survival probability as a function of circuit depth $L$ for 2-qubit randomized benchmarking using the FWCX and LCCX gate. At every depth, 50 random circuits are averaged over. The two simulation methods presented use the same 10 independent noise realizations using noise model NM1. For the subspace MC simulations, 5 Monte Carlo trials are performed per noise realizations. Error bars are $2\sigma$ confidence intervals generated by performing a bootstrap over the 10 noise realizations.}
   \label{fig:2QRB}
\end{figure}
\begin{figure*}[htbp] 
   \centering
   \includegraphics[width=7.in]{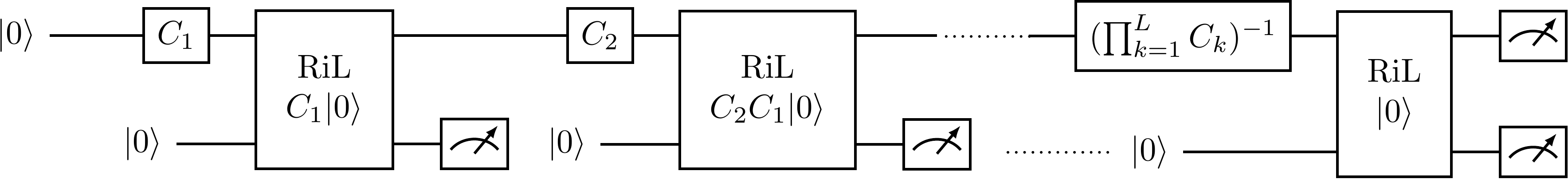} 
   \caption{\textbf{RiL gadget benchmarking protocol.} A depth $L$ 1-qubit RB circuit with RiL gates interleaved. Each RiL gate is chosen such that the reset state matches the ideal state at that point in the circuit.}
   \label{fig:RBRiL}
\end{figure*}
\begin{figure}[htbp] 
   \centering
   \includegraphics[width=3.25in]{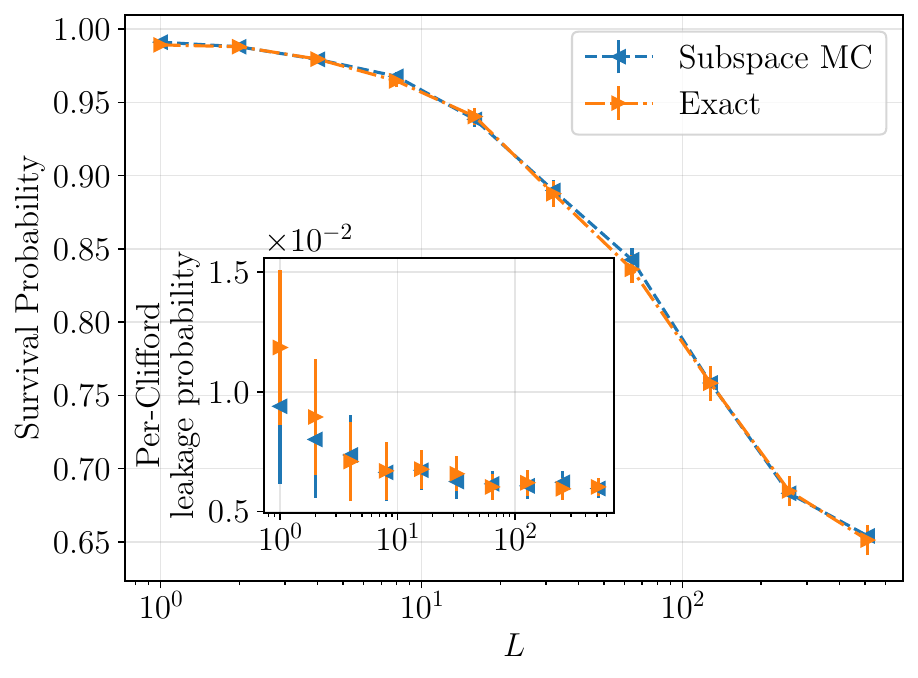} 
   \caption{\textbf{Simulation of the RiL gadget benchmarking protocol.} Average survival probability as a function of circuit depth $L$ for the RiL gadget benchmarking protocol illustrated in Fig.~\ref{fig:RBRiL}. At every depth, 25 random circuits are averaged over. The two simulation methods presented use the same 100 independent noise realizations. For the subspace MC simulations, 10 Monte Carlo trials are performed per noise realization. Error bars are $2\sigma$ confidence intervals generated by performing a bootstrap over the 100 noise realizations using noise model NM1. {The inset shows per-Clifford leakage probability as a function of circuit depth $L$. As expected, the leakage probability approaches a constant, corresponding to an asymptotic per-Clifford leakage probability of $6.1\times 10^{-3} \pm 2 \times 10^{-4}$ in this example.}}
   \label{fig:RBRiLResults}
\end{figure}
\subsubsection{2-qubit repeated CNOT} \label{sec:2qFWCX}
We now present a pathological case, where preserving the coherence between the leakage states and the computational states between operations is crucial. We consider the FWCX gate, which generates a coherent superposition of states in both the leakage and computational subspaces when acting on a target qubit that is in a leaked state even in the absence of noise (as illustrated in Eq.~\eqref{eqt:LeakagePropagaion}). A pair of FWCX gates (see Fig.~\ref{fig:RepeatedCXCircuit}) generates an identity operation (up to a global phase) irrespective of its input states.  Therefore, in order for a pair of FWCX gates to map the input state in Eq.~\eqref{eqt:LeakagePropagaion} back to itself, we must preserve the coherences generated by the first FWCX gate. However, our subspace projection decoheres the superposition of leakage and computational states, so it fails to realize the correct operation, even in the absence of noise and with or without randomized compiling. We demonstrate this failure in Figs.~\ref{fig:RepeatedFWCXResults} and \ref{fig:RepeatedRCFWCXResults}.

While this example was chosen to highlight how our subspace projection approach may fail, we believe this failure mechanism is not realized often enough in quantum circuits of interest to be of serious concern. In order to justify this, we show 2-qubit randomized benchmarking~\cite{Magesan2011} results where state preparation errors are introduced as the only source of error. We choose a state preparation error model where with probability $1-p$ the state $\ket{S_0} = \frac{1}{\sqrt{2}} \left( \ket{\uparrow \downarrow} - \ket{\downarrow \uparrow } \right)$ on two dots is prepared, with probability $p/2$ the state $\ket{T_0}= \frac{1}{\sqrt{2}} \left( \ket{\uparrow \downarrow} + \ket{\downarrow \uparrow } \right)$ is prepared, and with probability $p/2$ the state $\ket{T_-} = \ket{\downarrow \downarrow}$, such that the probability of preparing the three-dot states is:
\begin{eqnarray}
\Pr(\ket{0}_{-1/2}) &=& \frac{1}{2} \left( 1 - p \right) \ , \ 
\Pr(\ket{0}_{1/2}) = \frac{1}{2} \left( 1 - p \right) \ , \nonumber \\ 
\Pr(\ket{1}_{-1/2}) &=& \frac{1}{4} p \ , \ 
\Pr(\ket{1}_{1/2}) = \frac{1}{12} p \ , \nonumber \\
\Pr(\ket{L}_{-3/2}) &=& \frac{1}{4} p \ , \ x
\Pr(\ket{L}_{-1/2}) = \frac{1}{4} p \ , \nonumber \\
\Pr(\ket{L}_{1/2}) &=& \frac{1}{6} p \ .
\end{eqnarray}
Therefore, with probability $2p/3$ a leaked state is prepared on any given EO qubit.  In Fig.~\ref{fig:2QRB_stateprep}, we show the survival probability, i.e. the probability of measuring 0 for both qubits, in this case, and we see excellent agreement between the exact and Subspace MC simulations for a range of $p$ values.

It is worth emphasizing that the randomized compiling and benchmarking performed here is nominally meant to twirl the noise in the computational subspace, so it may be surprising to find that it works when including leaked states as well.  We believe this is related to the fact that each Clifford gate (as defined in Ref.~\cite{Andrews2019}) acting on a leaked state preserves the leaked state but imparts a phase that depends on the Clifford gate. Therefore, in the multi-qubit setup, this results in a randomization of the relative phase associated with a leaked state.

\subsubsection{2-qubit randomized benchmarking} \label{sec:2qRB}
%
Given the agreement observed for the randomized compiling of the repeated SWAP circuit, we can expect good agreement for the survival probability of 2-qubit randomized benchmarking~\cite{Magesan2011}. We use the decomposition of 2-qubit Cliffords into 1-qubit Cliffords, SWAP, CNOT described in Ref.~\cite{Corcoles2013}, and we show results using both the FWCX and the LCCX gate. (We note that an implementation of the iSWAP gate for exchange-only qubits~\cite{Madzik2025,Chadwick2025} could be used as well for a more efficient decomposition.) We see excellent agreement between the exact and Subspace MC simulations as shown in Fig.~\ref{fig:2QRB}. We note that the survival probability drops below $1/4$ and converges towards $1/12$. This is the expected value for our simulations because we are only considering 2-qubit initial states with $s_z \in \left\{-1, 0, 1 \right\}$ with probability 1/4, 1/2, 1/4 respectively. If the state converges at large depths to the maximally mixed state in each of these subspaces, which have dimension 15, 20, and 15 respectively, then the probability of measuring outcome 00 at large depth is given by $\frac{1}{4} \left(\frac{1}{4} \frac{4}{15} + \frac{1}{2} \frac{8}{20} + \frac{1}{4} \frac{4}{15}  \right)= 1/12$.

\subsubsection{1-qubit randomized benchmarking with RiL}
%
The previous examples do not directly probe leakage, since a leaked state is measured as a computational 1 state. One way to study leakage is using a RiL gadget~\cite{Langrock2020}, and we can construct different variants of the gadget such that each variant resets the qubit to a different computational state. We can use this to construct a circuit that resembles a 1-qubit RB sequence, where after every 1-qubit Clifford operation we have a RiL gadget that resets to the ideal state at that point in the RB circuit. This is illustrated in Fig.~\ref{fig:RBRiL}. The advantage of this construction is that we are not restricted to post-selecting on the subset of the RB sequences where no leakage is detected when performing an RB analysis. In Fig.~\ref{fig:RBRiLResults}, we show excellent agreement between the exact and Subspace MC simulations for both the survival probability and the per-Clifford leakage probability (inset). 
%

%

%
\subsection{Larger scale simulations}
%
The examples from the previous sections have helped us identify when we can use the Subspace MC method with confidence, for example in circuits that twirl the noise. We now proceed to simulate circuits at larger scales where exact simulations become prohibitive. 

\subsubsection{2-qubit randomized benchmarking with RiL} \label{sec:2qRBwithRiL}
%
\begin{figure}[htbp] 
   \centering
    \subfigure[]{\includegraphics[width=2.75in]{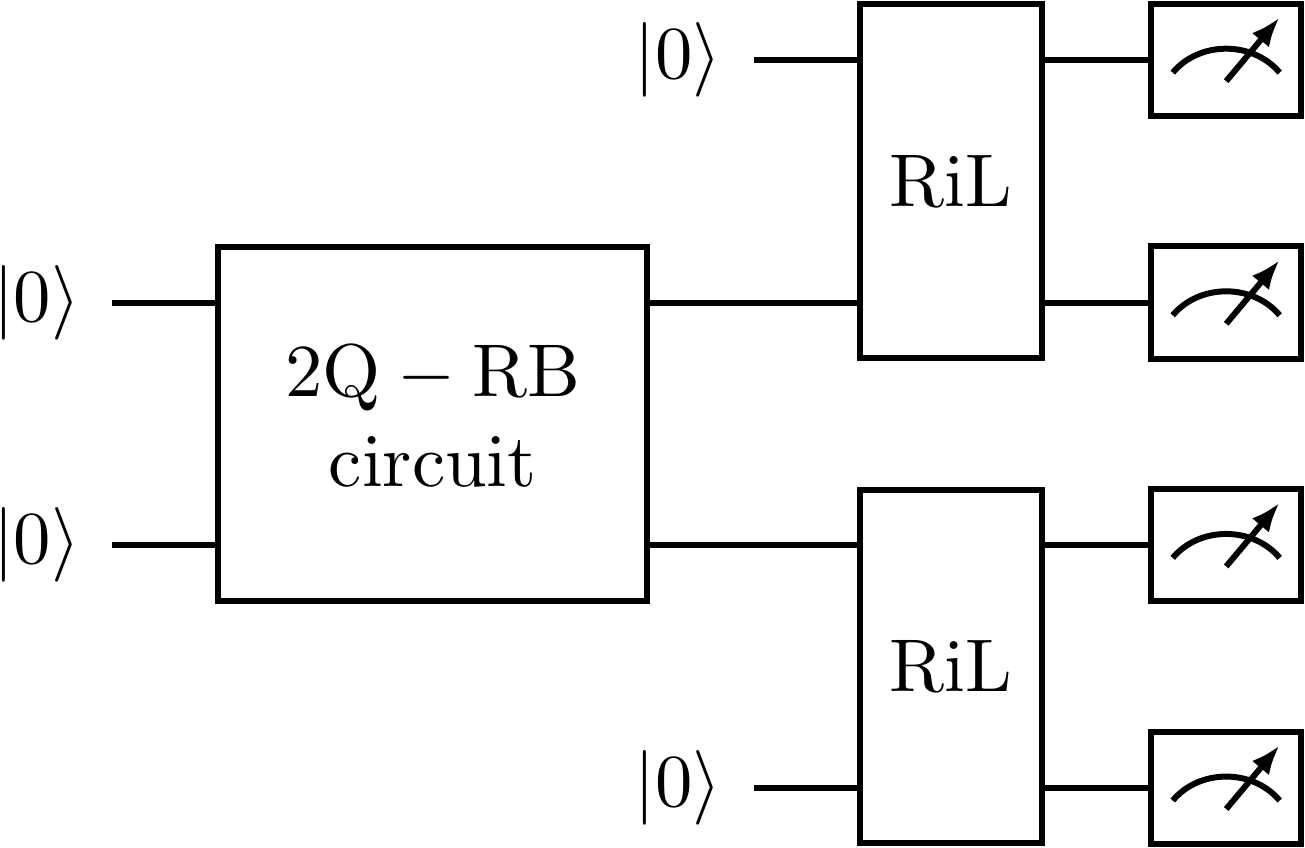} \label{fig:2qRBwithRiL}}
   \subfigure[]{\includegraphics[width=3.25in]{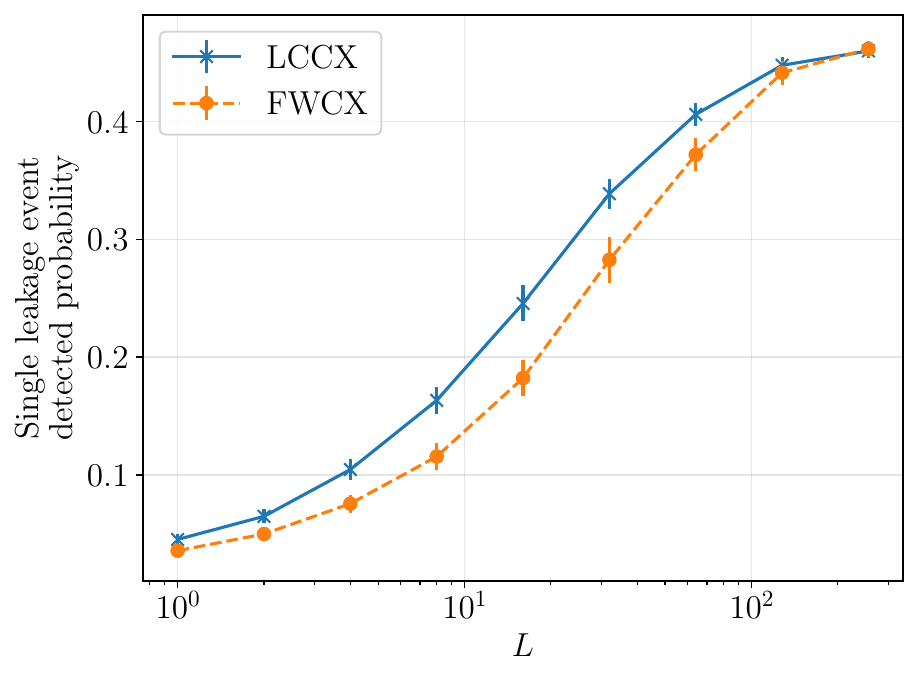} \label{fig:2QRBwithRiLsingle}}
   \subfigure[]{\includegraphics[width=3.25in]{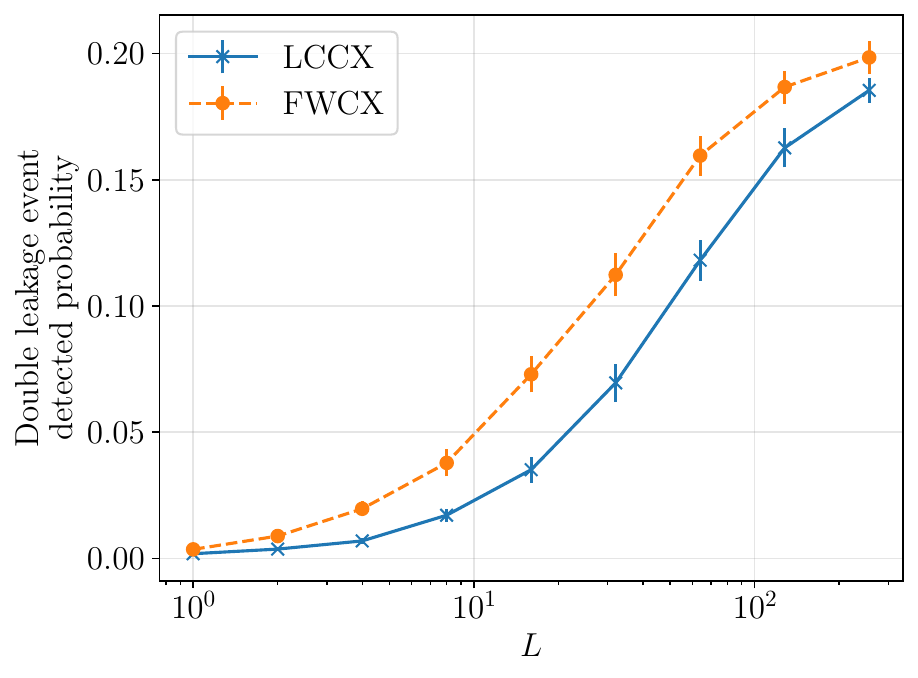} \label{fig:2QRBwithRiLdouble}}
   \caption{\textbf{Two-qubit RB with terminal leakage detection.} Results as a function of circuit depth $L$ for 2-qubit randomized benchmarking with RiL gadgets as depicted in (a). At every depth, 50 random circuits are averaged over. (b) The probability of only one of the two RiL gadgets detecting a leakage event. (c) The probability of both RiL gadgets detecting a leakage event. The results are averaged over 40 independent noise realizations using noise model NM1, and 5 Monte Carlo trials are performed per noise realizations. Error bars are $2\sigma$ confidence intervals generated by performing a bootstrap over the 40 noise realizations.}
   \label{fig:2QRBwithRiL}
\end{figure}
We consider a slight modification of the 2-qubit RB circuits studied in Sec.~\ref{sec:2qRB} in which we introduce an RiL gadget before the terminating measurement. The goal of this extension is to contrast the behavior of the LCCX and FWCX gates. Since each RiL gadget requires one ancilla qubit, the simulation involves effectively four EO qubits as illustrated in Fig.~\ref{fig:2qRBwithRiL}. Since the ancilla qubits only participate at the end of the RB circuit, we will further assume that the ancilla qubits are reset right before the implementation of the RiL gadget, so this extension is minimal in terms of its additional simulation demands. To induce more leakage in our circuits, we insert after every 2-qubit Clifford operation an idle time of 400ns. This choice is such that after post-selecting on measurement outcomes with no leakage, the survival probability is very similar for circuits using either type of CNOT gate (shown in Appendix~\ref{app:MoreResults}). In contrast to Fig.~\ref{fig:2QRB}, the post-selected survival probability converges to a value consistent with 1/4, which is what we would expect for a maximally-mixed state of computational states only.

We can investigate the role of leakage spreading by counting the number of circuits at a given depth where only a single leakage event is detected (Fig.~\ref{fig:2QRBwithRiLsingle}) versus two leakage events being detected (Fig.~\ref{fig:2QRBwithRiLdouble}). We find that the circuits using the FWCX gate have fewer single-leakage events compared to circuits using the LCCX gate, which can be attributed to these circuits being shorter in duration and hence less susceptible to magnetic noise. However, the circuits using the FWCX gate have more double-leakage events compared to circuits using the LCCX gate, indicating that leakage is more likely to spread using the FWCX gate than the LCCX gate, as expected.

\subsubsection{Stabilizing Bell State Circuits with RiLs} \label{sec:ParityCheckwithRiL}
%
\begin{figure*}[htpb]
   \centering
   \includegraphics[width=4.75in]{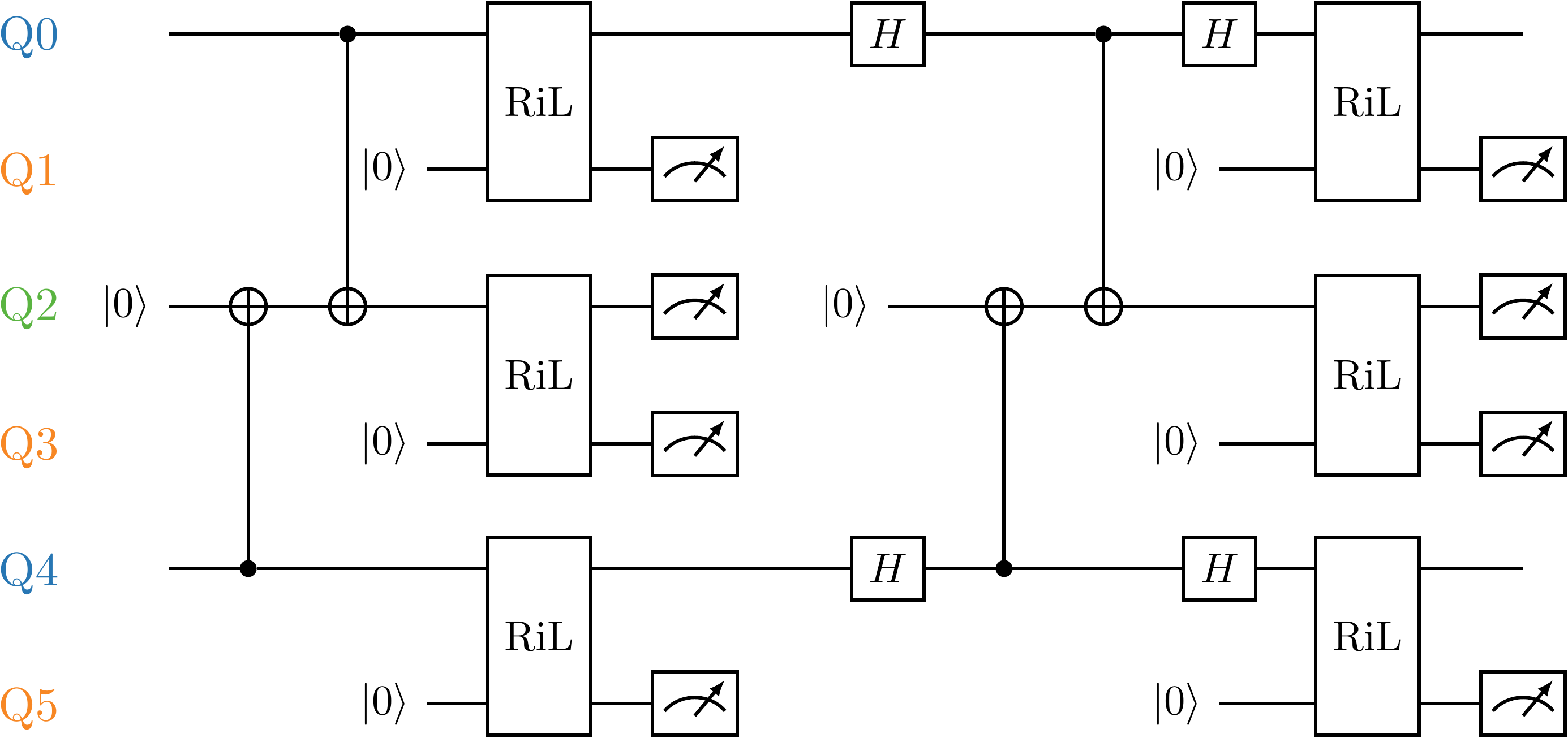} 
   \caption{\textbf{Bell state stabilization circuit using six EO qubits.} A single round of Z-parity and X-parity check circuits with RiL gadgets is applied to the data qubits (Q0 and Q4) and the measurement ancilla qubit (Q2). Each RiL gadget uses its own ancilla qubit (Q1, Q3, and Q5). The measurement and RiL gadget ancilla qubits are measured after each parity circuit.} 
   \label{fig:ParityCheckwithRiL}
\end{figure*}

We now consider the stabilization of a Bell state by repeated applications of Z-parity and X-parity check circuits, where we also apply a RiL gadget on the data and ancilla qubit after every parity check. (A similar realization of this experiment was done in superconducting qubits~\cite{Bultink2020} but without a RiL gadget.) This circuit involves six EO qubits, as illustrated in Fig.~\ref{fig:ParityCheckwithRiL}. The Z-parity circuit outputs 0 (1) if the state has even (odd) parity in the computational basis. Similarly, the X-parity circuit outputs 0 (1) if the state has even (odd) parity in the $x$ basis. The paired application of these measurements ideally projects the state onto the corresponding Bell state:
\bes
\begin{align}
\text{Outcomes: } 0_Z 0_X \to \ket{\Phi_0} = \frac{1}{\sqrt{2}} \left( \ket{00} + \ket{11} \right) \ , \\
\text{Outcomes: } 0_Z 1_X \to \ket{\Phi_1} = \frac{1}{\sqrt{2}} \left( \ket{00} - \ket{11} \right) \ , \\
\text{Outcomes: } 1_Z 0_X \to \ket{\Phi_2} = \frac{1}{\sqrt{2}} \left( \ket{01} + \ket{10} \right) \ , \\
\text{Outcomes: } 1_Z 1_X  \to \ket{\Phi_3} = \frac{1}{\sqrt{2}} \left( \ket{01} - \ket{10} \right) \ . 
\end{align}
\ees

We initialize the two data qubits in the Bell state $\ket{\Phi_0}$, which has even parity for both Z and X parity circuits. If the noise is sufficiently low, repeated applications of the Z- and X-parity checks stabilizes this state. However, a change in the Z-parity measurement signals that a single bit flip error on the state has been detected (for example taking $\ket{\Phi_0} \to \ket{\Phi_2}$). Similarly, a change in the X-parity measurement signals that a single phase flip error has been detected (for example taking $\ket{\Phi_0} \to \ket{\Phi_1}$). When such events are detected, the state is projected to the corresponding Bell state. By tracking the Z- and X-parity measurement outcomes, we can track the state of the system. This is referred to as Pauli frame updating~\cite{Knill2005,Kelly2015}.

The discussion above does not take into account what happens when the data and ancilla qubit leak out of the computational subspace. Leaked states behave differently from $\ket{1}$ under logical operations but are measured as $\ket{1}$, so they can result in incorrect state tracking and hence prevent the stabilization of the Bell state. The inclusion of RiL gadgets after every parity check means that qubits can be returned to the computational subspace when they leak, and the following round of parity measurements should return the pair of qubits to one of the four Bell states. This means that the inclusion of the RiL should allow for Bell state stabilization for sufficiently low error rates.

To ensure the validity of our simulation, we replace every CNOT in the circuit by a Clifford randomized compiled version. While we could further optimize the circuit by absorbing the neighboring pairs of single qubit Cliffords that appear on the ancilla qubit (Q2) between the two CNOTs, we do not do this.

\begin{figure}[htbp] 
   \centering
   \subfigure[]{\includegraphics[width=3.25in]{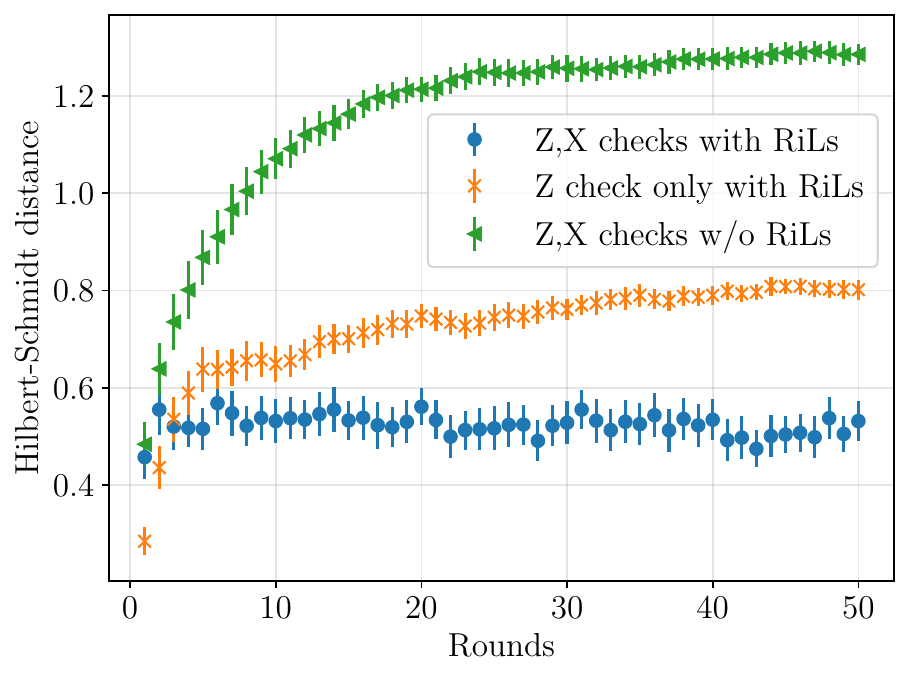}\label{fig:HSDistanceChecks}}
   \subfigure[]{\includegraphics[width=3.25in]{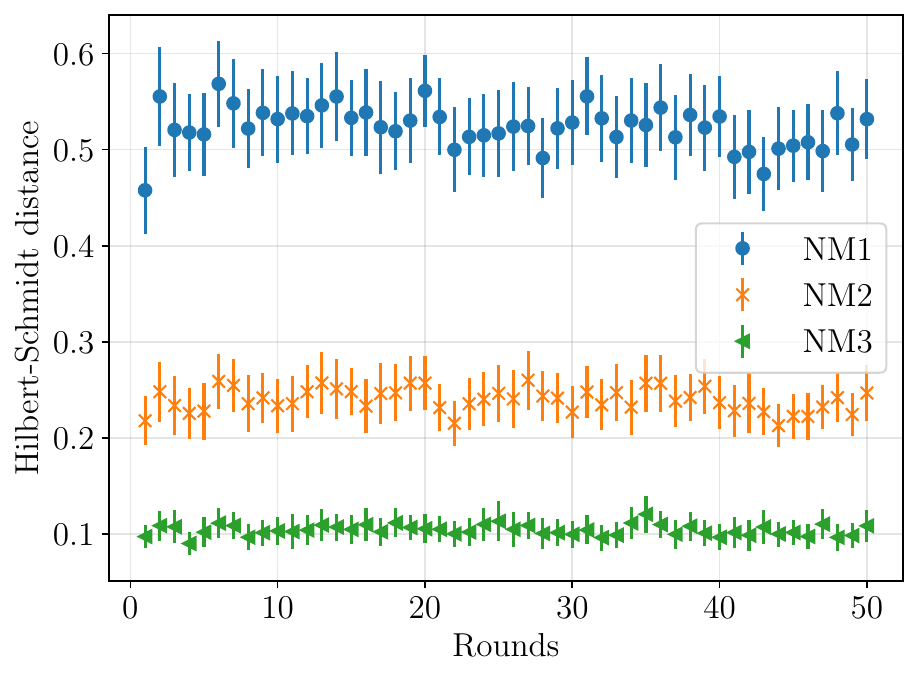} \label{fig:HSDistanceNoise}}
   \caption{\textbf{Simulated Bell state stabilization.} Hilbert-Schmidt distance of the simulated state with the expected Bell state based on the parity measurements. (a) Results for the three distinct cases: (1) both Z- and X-parity measurements are performed with RiLs, (2) only Z-parity measurements are performed with RiLs, and (3) both Z- and X-parity measurements are performed but without RiLs. These simulations use noise model NM1. (b) Results for case (1) with the three noise models NM1, NM2, and NM3. The results in (a) and (b) use the FWCX gate and are averaged over 100 independent noise realizations, and 10 Monte Carlo trials are performed per noise realizations. Error bars are $2\sigma$ confidence intervals generated by performing a bootstrap over the 100 noise realizations.}
   \label{fig:HSDistance}
\end{figure}

We confirm these expectations in Fig.~\ref{fig:HSDistance}, where we check Bell state stabilization by showing the behavior of the Hilbert-Schmidt distance, 
\begin{eqnarray}
\mathrm{HS}\left( \hat{\rho}_{s_{z}^{(1)}, \dots, s_{z}^{(n)}},  \ketbra{\Phi_k}{\Phi_k}\right) &=& \nonumber \\
&& \hspace{-3cm}  \sqrt{\Tr \left[ \left( \hat{\rho}_{s_{z}^{(1)}, \dots, s_{z}^{(n)}}  - \ketbra{\Phi_k}{\Phi_k}  \right)^2 \right]} \ , 
\end{eqnarray}
between the simulation state $\hat{\rho}_{s_{z}^{(1)}, \dots, s_{z}^{(n)}}$ and the predicted Bell state $\ket{\Phi_k}$ based on the parity measurements as a function of measurement rounds for a few noise models. These different noise models, labeled NM1, NM2, and NM3, have decreasing magnetic and voltage error rates (increasing $T_2^\ast$ value), and details are given in Tables~\ref{table:magneticT2} and \ref{table:voltageT2}. 
In Fig.~\ref{fig:HSDistanceChecks}, we check Bell state stabilization for different cases: (1) both Z- and X-parity measurements are performed with RiLs, (2) only Z-parity measurements are performed with RiLs, and (3) both Z- and X-parity measurements are performed but without RiLs. For cases (2) and (3), they both result in increasing Hilbert-Schmidt distance with increasing rounds so are not able to stabilize the state. Only case (1) results in a constant Hilbert-Schmidt distance with increasing rounds and hence stabilize the state. In Fig.~\ref{fig:HSDistanceNoise}, we show how the value of the Hilbert-Schmidt distance of the stabilized state decreases with decreasing error rates.
\begin{figure}[htbp] 
   \centering
   \subfigure[FWCX]{\includegraphics[width=3.25in]{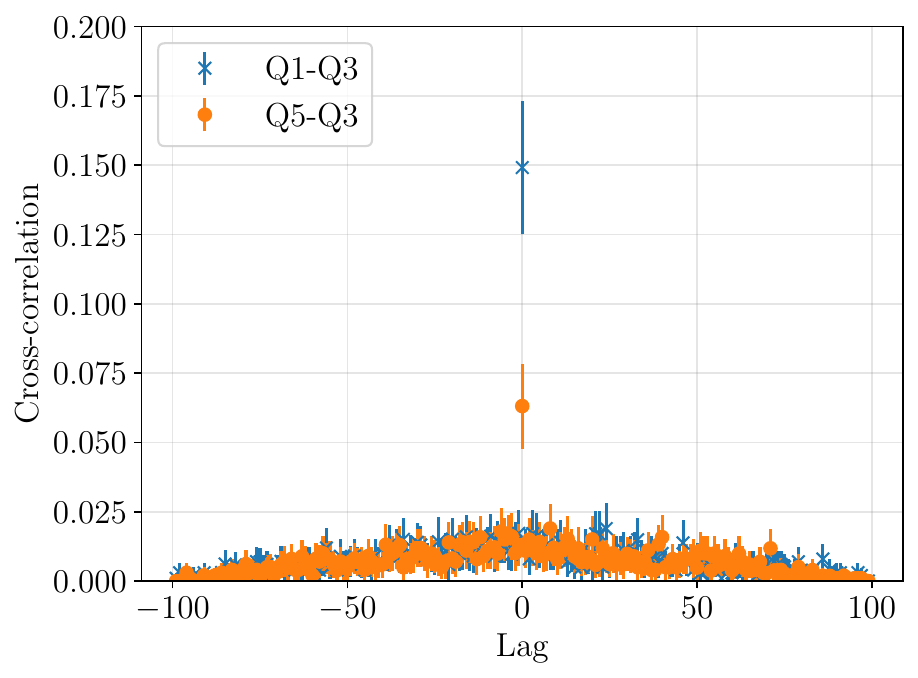}\label{fig:CrosscorrelationFW}}
   \subfigure[LCCX]{\includegraphics[width=3.25in]{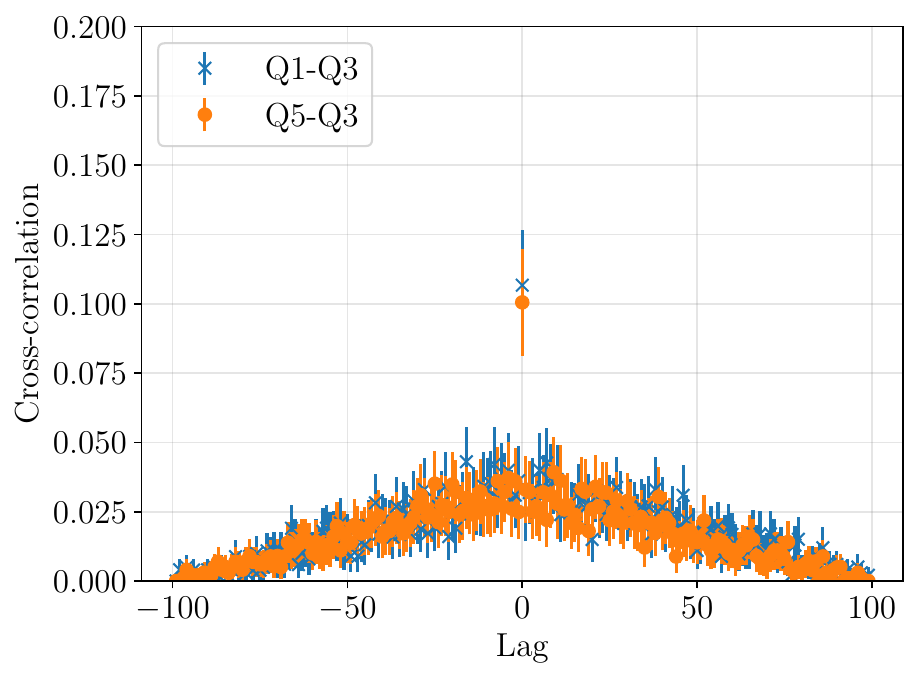} \label{fig:CrosscorrelationLC}}
   \caption{\textbf{Cross-correlation of the RiL measurement outcomes of the data qubits (Q1, Q5) with the measurement ancilla (Q3).}  Cross-correlation calculated using signal.correlate method in SciPy 1.16.3~\cite{SciPy-NMeth2020}. Results use noise model NM1. The results in (a) use the FWCX gate, and the results in (b) use the LCCX gate. Results are averaged over 100 independent noise realizations, and 10 Monte Carlo trials are performed per noise realization. Error bars are $2\sigma$ confidence intervals generated by performing a bootstrap over the 100 noise realizations.}
   \label{fig:CrossCorrelation}
\end{figure}
A comparison between simulations with the three noise models NM1, NM2, and NM3 that use the shorter FWCX gate exhibit lower Hilbert-Schmidt distances for the stabilized state compared to simulations that use the longer but leakage controlled LCCX gate. We refer to these two sets of circuits as FWCX and LCCX circuits. This indicates that for the noise models we study the physical circuit length remains the indicator of performance. Nevertheless, we do see differences in the correlations of leakage events detected. When studying the correlation of leakage detection events between the data qubits (the measurement outcomes of Q1 and Q5) and the measurement ancilla qubit (the measurement outcome of Q3), we find a statistically significant correlation for same-time detection for both FWCX and LCCX circuits, as shown in Fig.~\ref{fig:CrossCorrelation}. For the FWCX circuits (Fig.~\ref{fig:CrosscorrelationFW}), we observe that there is a stronger correlation between Q1 and Q3 compared to Q5 and Q3, meaning that there is a stronger correlation with the second data qubit operated on rather than the first. This difference in correlation is absent for the LCCX circuits (Fig.~\ref{fig:CrosscorrelationLC}), which suggests that the cause of the difference in correlation between Q1-Q3 and Q5-Q3 is leakage spreading from the measurement ancilla but only after operating on the first data qubit. In both cases, the triangular behavior of the cross-correlation away from a lag of zero is consistent with independent random leakage detection events.

\begin{figure}[t] 
   \centering
   \subfigure[FWCX]{\includegraphics[width=3.25in]{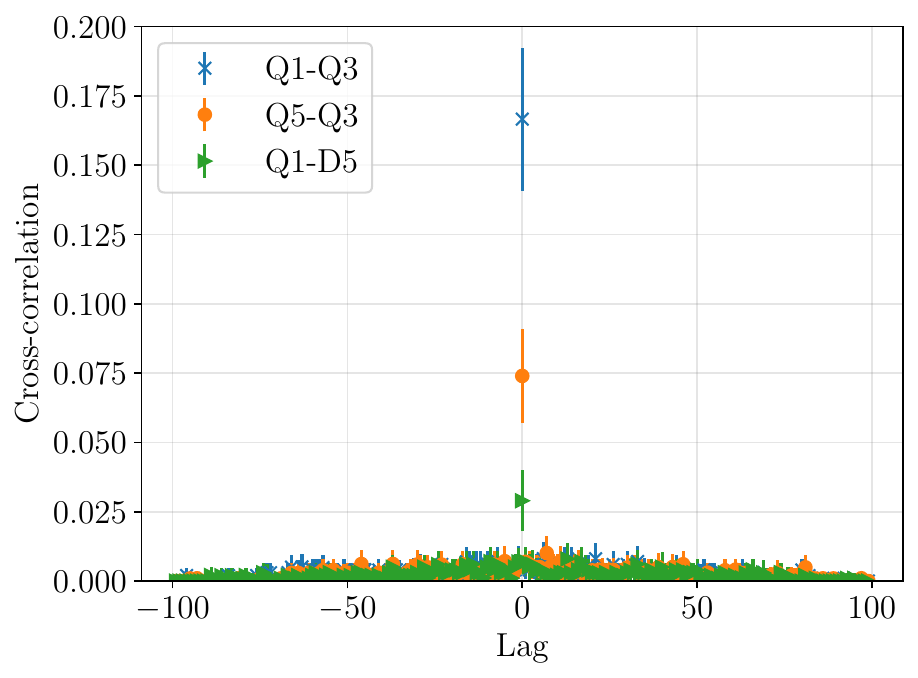}\label{fig:CrosscorrelationFWNoIdle}}
   \subfigure[LCCX]{\includegraphics[width=3.25in]{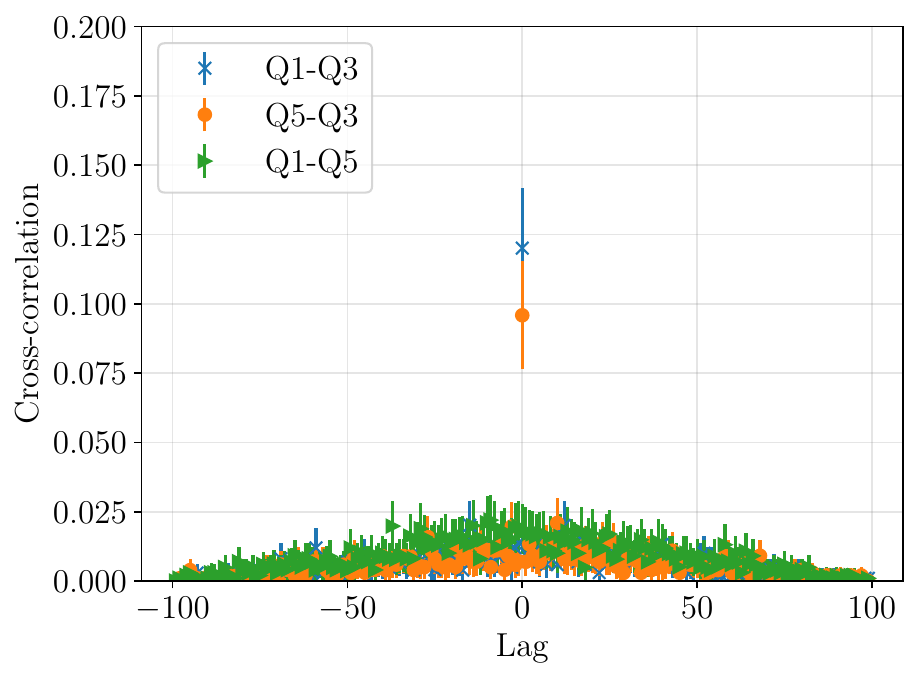} \label{fig:CrosscorrelationLCNoIdle}}
   \caption{\textbf{Cross-correlation of the RiL measurement outcomes of the data qubits (Q1, Q5) and the measurement ancilla (Q3) when the data qubits are not affected by noise during idling.} Cross-correlation calculated using signal.correlate method in SciPy 1.16.3~\cite{SciPy-NMeth2020}. Results use noise model NM1. The results in (a) use the FWCX gate, and the results in (b) use the LCCX gate. Results are averaged over 100 independent noise realizations, and 10 Monte Carlo trials are performed per noise realization. Error bars are $2\sigma$ confidence intervals generated by performing a bootstrap over the 100 noise realizations.}
   \label{fig:CrossCorrelationNoIdle}
\end{figure}
We can ask why there is any correlation between the leakage detection events for the LCCX circuits, since the gate ideally should not cause leakage to spread. If we turn off both the magnetic and the voltage noise only during the implementation of the LCCX gate, we find that the cross-correlation effectively vanishes. This indicates that the noise (both magnetic and voltage) during the implementation of the gate causes these 2-qubit correlations. 

However, we do not observe a similarly strong statistically significant correlation between the two data qubit measurements. These correlations are weaker and are drowned out by the effect of magnetic noise when the data qubits are idling. In Fig.~\ref{fig:CrossCorrelationNoIdle}, we show the cross-correlation when we artificially eliminate the noise when idling, and we now observe a small but statistically significant cross-correlation between the data qubits when using the FWCX gates (Fig.~\ref{fig:CrosscorrelationFWNoIdle}) but not when using the LCCX gates (Fig.~\ref{fig:CrosscorrelationLCNoIdle}).

%
\section{Conclusions} \label{sec:conclusions}
%
We have presented and demonstrated an approach for simulating the dynamics of multi-EO qubits in the presence of magnetic noise along the global field direction and voltage noise. Motivated by the observation that noiseless qubit operations with non-leakage states preserve the individual qubit's {spin projection quantum number along the $z$-axis}, we project each qubit onto a subspace with a {definite spin projection quantum number} after every logical operation. This allows us to express the quantum state of $n$ EO qubits in terms of a system of $n$ qutrits with an additional label for each qubit storing its $s_z$ value. We find that this approximation follows the exact dynamics for circuits that convert coherent noise to stochastic noise via twirling~\cite{Bennett1996a,Bennett1996b}, which can be readily realized through randomized compiling~\cite{Wallman2016}. 

We emphasize that the approximation made by our subspace projection happens on the 3- or 6-spin output state generated by the implementation of the 1- or 2-qubit logical operation respectively and not during the implementation of the operation. Our simulation remains a spin-level simulator during these operations, so a faithful representation of the effect of magnetic and voltage noise on the logical operation can be realized. It is the state between these operations that is being truncated to achieve a favorable scaling.

We highlight the capabilities of this approach by simulating circuits with 4 and 6 EO-qubits, where we study the role of leakage-controlled CNOT gates to suppress correlated leakage events in 2-qubit randomized benchmarking and Bell state stabilization. In these simulations, we did not include state preparation or measurement errors, nor did we include a finite time for measurements, all of which can be easily included and would be required when performing comparisons to experiments.

{In demonstrating the capabilities of the Subspace MC approach, we developed a new protocol that may allow for improved characterization of the RiL gadget. In addition, in order to understand our observed statistics of leakage detection outcomes in the Bell state stabilization circuit, we explored selectively turning off noise during dynamics to probe when and where specific leakage errors occur. We suggest that this approach, generalizing our previous approach to error attribution \cite{Gaye2025}, may in the future provide insights into observed syndrome measurement outcomes for QEC circuits.}

We comment on the implications of our choice of model for the magnetic noise. By only considering magnetic noise along the global field direction, the subspace projection for a single qubit is not an approximation since the system and noise Hamiltonians preserve the {spin projection quantum number along the $z$-axis} of the qubit, so the first non-trivial test for the method occurs at two qubits. In the 2-qubit case, the system and noise Hamiltonians preserve the sum of the spin projection quantum numbers of the two qubits, which allow us to implement temporal coarse graining in subspaces rather than the full 6-spin Hilbert space of 2 EO-qubits. We can relax our assumption on the magnetic noise so that it has no axis restriction, and we do not expect this to change our conclusions for the validity of our method. However, the subspace projection would be an approximation even in the 1-qubit case, and the temporal coarse graining would need to be performed on the full space as opposed to subspaces.

While Subspace MC extends the size of simulations of EO qubits that can be performed, further order reduction will be needed to simulate even larger system sizes, such as those relevant for quantum error correction (QEC)~\cite{Shor1996,Aharonov1997,Knill1997,Knill1998,Knill2000,Terhal2005,Aharonov2006}. The success of the Subspace MC approach for noise-twirled circuits suggests it is already providing an effective description of the noise in terms of a time-dependent depolarizing channel or generalized Pauli noise channel, depending on the type of noise twirling, for our effective qutrit description. We expect incorporating further approximations to Subspace MC, such as subspace twirling~\cite{Marshall2025} that preserves incoherence between the computational subspace and leakage subspace, to be a fruitful direction for generating even more scalable reduced order noise models for QEC circuits using EO qubits while still capturing relevant features of the magnetic and voltage noise over multiple time scales. We leave this exploration for future work.
 
\begin{acknowledgements}
This work was performed, in part, at the Center for Integrated Nanotechnologies, an Office of Science User Facility operated for the U.S. Department of Energy (DOE) Office of Science.

This article has been co-authored by an employee of National Technology \& Engineering Solutions of Sandia, LLC under Contract No. DE-NA0003525 with the U.S. Department of Energy (DOE). The employee owns all right, title and interest in and to the article and is solely responsible for its contents. The United States Government retains and the publisher, by accepting the article for publication, acknowledges that the United States Government retains a non-exclusive, paid-up, irrevocable, world-wide license to publish or reproduce the published form of this article or allow others to do so, for United States Government purposes. The DOE will provide public access to these results of federally sponsored research in accordance with the \href{https://www.energy.gov/downloads/doe-public-access-plan.}{DOE Public Access Plan}.
\end{acknowledgements}
\bibliography{refs}

\appendix  
\section{Basis decomposition} \label{app:BasisDecomposition}
We give a decomposition of the states in Eq.~\eqref{eqt:EightBasisStates} in terms of the spin $\ket{\uparrow}, \ket{\downarrow}$ basis, where we define $\hat{S}_z \ket{\uparrow} = \frac{1}{2} \ket{\uparrow}$. If $\mathcal{H}_i$ is the 2-dimensional Hilbert space associated with the $i$-th spin, we assume the composite Hilbert space is given by the tensor product $\mathcal{H}_1 \otimes \mathcal{H}_2 \otimes \mathcal{H}_3$. The decomposition of the states $\left\{\ket{S,S_{12},S_z}\right\}$ is then given by: 
\bes \label{eqt:ExchangeBasis}
\begin{align}
\ket{\frac{1}{2}, 0, \frac{1}{2}} &= \frac{1}{\sqrt{2}} \left( \ket{\uparrow \downarrow \uparrow}  - \ket{ \downarrow \uparrow \uparrow}  \right) = \ket{S_0} \otimes \ket{\uparrow} \ , \\
\ket{\frac{1}{2}, 0, -\frac{1}{2}} &= \frac{1}{\sqrt{2}} \left( \ket{\uparrow \downarrow \downarrow}  - \ket{ \downarrow \uparrow \downarrow}  \right) = \ket{S_0} \otimes \ket{\downarrow} \ , \\
\ket{ \frac{1}{2} , 1 , \frac{1}{2} } &= \sqrt{\frac{2}{3}} \ket{ \uparrow \uparrow \downarrow} - \sqrt{ \frac{1}{6}} \left( \ket{\uparrow \downarrow \uparrow} + \ket{\downarrow \uparrow \uparrow} \right) \nonumber \\
& = \sqrt{\frac{1}{3}} \left( \sqrt{2} \ket{T_+} \ket{\downarrow} - \ket{T_0} \ket{\uparrow} \right) \ , \\
\ket{ \frac{1}{2} , 1 , -\frac{1}{2} } &= \sqrt{\frac{1}{6}} \left( \ket{ \uparrow \downarrow \downarrow} +   \ket{\downarrow \uparrow \downarrow} \right) - \sqrt{\frac{2}{3}} \ket{\downarrow \downarrow \uparrow} \nonumber \\
& = \sqrt{\frac{1}{3}} \left( \ket{T_0} \ket{\downarrow} - \sqrt{2} \ket{T_-} \ket{\uparrow} \right) \ , \\
\ket{\frac{3}{2}, 1, \frac{3}{2}} & =  \ket{\uparrow \uparrow \uparrow} = \ket{T_+} \ket{\uparrow} \ , \\
\ket{\frac{3}{2}, 1, \frac{1}{2}} &= \sqrt{\frac{1}{3}} \left( \ket{\uparrow \uparrow \downarrow} +  \ket{\uparrow  \downarrow \uparrow} +  \ket{  \downarrow \uparrow \uparrow} \right) \nonumber \\
&= \sqrt{\frac{1}{3}} \left( \ket{T_+} \ket{\downarrow} + \sqrt{2} \ket{T_0} \ket{\uparrow} \right) \ , \\
\ket{\frac{3}{2}, 1, -\frac{1}{2}} &= \sqrt{\frac{1}{3}} \left( \ket{\uparrow \downarrow \downarrow} +  \ket{\downarrow \uparrow \downarrow } +  \ket{  \downarrow \downarrow \uparrow} \right) \nonumber \\
&= \sqrt{\frac{1}{3}} \left( \sqrt{2} \ket{T_0} \ket{\downarrow} + \ket{T_-} \ket{\uparrow} \right) \ , \\
\ket{\frac{3}{2}, 1, -\frac{3}{2}} &= \ket{\downarrow \downarrow \downarrow} = \ket{T_-} \ket{\downarrow} \ ,
\end{align}
\ees
where the 2-spin singlet and triplet states are defined as:
\bes 
\begin{align}
\ket{S_0} &= \frac{1}{\sqrt{2}} \left( \ket{\uparrow \downarrow} - \ket{\downarrow \uparrow } \right) \,  \\
\ket{T_-} &= \ket{\downarrow \downarrow} \ ,  \\
\ket{T_0} &= \frac{1}{\sqrt{2}} \left( \ket{\uparrow \downarrow} + \ket{\downarrow \uparrow } \right) \ , \\
\ket{T_+} &= \ket{\uparrow \uparrow}  \ .
\end{align}
\ees

\section{Temporal Coarse Graining}  \label{app:TemporalCoarseGraining}
We briefly describe the temporal coarse graining method that we use. Details of this method can be found in Ref.~\cite{Albash2025}.
The noise Hamiltonian is assumed to take the form:
\beq
\hat{H}_{\mathrm{N}} = \sum_{\alpha} \eta_{\alpha}(t) \hat{B}_{\alpha}(t) \ ,
\eeq
where $\eta_{\alpha}(t)$ are a sum of independent Ornstein-Uhlenbeck (OU) processes~\cite{Uhlenbeck1930}. The unitary dynamics is expressed as a composition $\mathcal{U}_{\mathrm{I}} \cdot \tilde{\mathcal{U}}_{\mathrm{N}}$, where all the dependence on the noise Hamiltonian is in $\tilde{\mathcal{U}}_{\mathrm{N}}$.
Conditioned on a realization of the OU processes at predetermined time points $t_0$ and $t_1$(for example, at the beginning and end of the logical gates), the conditioned OU processes are expressed in terms of bridge stochastic processes that take value 0 at $t_0$ and $t_1$. We can perform an ensemble average over the bridge processes, leaving us with a quantum operation $\mathcal{U}_{\mathrm{I}} \cdot \mathcal{E}$ describing the evolution from $t_0$ to $t_1$ that depends only on the realization of the OU processes at these boundary time points.
\section{Magnetic noise} 
%
We give implementation details of the magnetic noise described in Eq.~\eqref{eqt:MagneticNoise}. Each $\delta h_i(t)$ is a sum of 6 OU processes with frequencies distributed log-linearly~\cite{Kaulakys2005} between 1 mHz and 100 kHz. Each OU process has a single-sided power spectral density (PSD) of the form:
\beq \label{eqt:OUpsd}
S(f) = \frac{1}{\pi} \frac{p f_0}{f_0^2 + f^2} \ ,
\eeq
where $f_0$ is the frequency of the OU process and $p$ is the same for all the OU processes. In order to achieve a free induction decay with a desired $T_2^\ast$ value, we choose $p$ values according to Table~\ref{table:magneticT2}. Further details about OU processes relevant to this construction can be found in Ref.~\cite{Albash2025}.
\begin{table}[!htbp]
\centering
\begin{tblr}{
  colspec = {|c|c|c|},
}
\hline
Noise model label & $T_2^\ast$ ($\mu$s)     &    $p/h^2$  (MHz$^2$) \\
  \hline[2pt]
\cline{1-4}
NM1 & $3.5$ & $7.45654 \times 10^{-4}$ \\
NM2 & $7$ &   $1.90008 \times 10^{-4}$\\
NM3 & $14$ & $4.84165 \times 10^{-5}$\\
\hline
\end{tblr}
\caption{Magnetic noise OU parameter $p$ to achieve a desired free induction decay $T_2^\ast$ value.} \label{table:magneticT2}
\end{table}
\section{Voltage noise}  \label{app:ChargeNoise}
%
We model the exchange interaction between two spins as a function of applied voltage $V(t)$ with the Hamiltonian:
\begin{eqnarray}
\hat{H}_{\mathrm{EX}} &=& J_0 e^{\frac{V(t) + \delta V(t)}{\mathcal{I}}} \vec{\hat{S}}_1 \cdot \vec{\hat{S}}_2 \nonumber \\
& =& \left(    J_0 e^{\frac{V(t)}{\mathcal{I}}} +   J_0 e^{\frac{V(t)}{\mathcal{I}}} \left( e^{\frac{ \delta V(t)}{\mathcal{I}}} - 1 \right) \right)   \vec{\hat{S}}_1 \cdot \vec{\hat{S}}_2 \nonumber \\
&=& \left( J(t) + J(t) \left( e^{\frac{ \delta V(t)}{\mathcal{I}}} - 1 \right) \right)   \vec{\hat{S}}_1 \cdot \vec{\hat{S}}_2 \ ,
\end{eqnarray}
where $\mathcal{I}$ is the insensitivity and $\delta V(t)$ is the charge noise given by a stochastic process given by a sum of OU processes. To fit into the formalism described in Appendix~\ref{app:TemporalCoarseGraining}, we take $\eta =  \left( e^{\frac{ \delta V(t)}{\mathcal{I}}} - 1 \right) $ and $\hat{B}(t) = J(t) \vec{\hat{S}}_1 \cdot \vec{\hat{S}}_2$, where $\eta$ is a stochastic process that depends on an exponential of OU processes. While we can expand the exponential such that $\eta \approx \delta V / \mathcal{I}$ to leading order and have $\eta$ be a sum of OU processes as for the case of magnetic noise, we can make progress with the exponential of a sum of OU processes. This is a departure from the derivation presented in Ref.~\cite{Albash2025}, so we present it here for completeness. 

We denote $\delta V(t)/\mathcal{I} = \sum_{n} X_n(t)$, where $X_n(t)$ are independent OU processes, satisfying the stochastic differential equation:
\beq
d X_n(t) = \gamma_n \left( \mu_n - X_n(t) dt \right)+ \sigma dW(t) \ ,
\eeq
For simplicity we will take $\mu_n = 0$.
We define $Y(t) = \exp \left( \sum_n X_n(t)\right)$ such that $\eta(t) = Y(t) - 1$. When we condition on a realization of the noise at two time points, $\partial_n \equiv \left(X_n(t_{k-1}) = x_{n,k-1}, X_n(t_{k}) = x_{n,k} \right)$, we can express the conditioned OU process as:
\begin{eqnarray} \label{eqt:conditionedOUprocess}
\left( X_n(t) | \partial_n \right) &=& \mathbb{E} \left[ X_n(t) | \partial_n \right]  + X_n'(t) \ , 
\end{eqnarray}
where $X_n'(t)$ is a bridge process that takes value zero at its boundaries. We refer to this as a zero-boundary bridge process. If we condition all the OU processes, $\partial \equiv \left\{ \partial_n \right\}_n$, the conditioned $Y$ process is given by:
\begin{eqnarray} \label{eqt:conditionedExpOUprocess}
\left( Y(t) | \partial \right) &=& \exp \left( \sum_n \mathbb{E} \left[ X_n(t) | \partial_n \right]  + X_n'(t)\right) \ . 
\end{eqnarray}
We want to express this conditioned process in a way similar to how we expressed the conditioned OU process in Eq.~\eqref{eqt:conditionedOUprocess}. Using the cumulant expansion and the independence of the OU processes (recall that the bridge processes have zero mean), the expectation value of Eq.~\eqref{eqt:conditionedExpOUprocess} is given by:
\begin{eqnarray}
\mathbb{E} \left[ Y(t) | \partial  \right]  &=&  \exp \left( \sum_n \mathbb{E} \left[ X_n(t) | \partial_n \right] \right) \nonumber \\
&& \times \mathbb{E} \left[ \exp \left(  \sum_n X'_n(t)   \right) \right] \nonumber \\
&=&  \exp \left( \sum_n \mathbb{E} \left[ X_n(t) | \partial_n \right] \right) \nonumber \\
&& \times \exp \left(  \frac{1}{2} \sum_n \mathrm{Var} \left[  X'_n(t) \right]\right) \ .
\end{eqnarray}
We therefore write:
\beq
\left( Y(t) | \partial \right) = \mathbb{E} \left[ Y(t) | \partial  \right]  + \left( Y'(t) | \partial \right) \ ,
\eeq
where $\left( Y'(t) | \partial \right)$ has zero mean and is given by
\begin{eqnarray}
\left( Y'(t) | \partial \right) &=& \exp \left( \sum_n \mathbb{E} \left[ \left(X_n(t) | \partial_n  \right) \right] \right) \nonumber \\
&& \hspace{-1.5cm} \times \left(  \exp \left(   \sum_n X'_n(t)  \right) -  \exp \left(  \frac{1}{2} \sum_n \mathrm{Var} \left[ X'_n(t)  \right] \right) \right) \ . \nonumber \\
\end{eqnarray}
Expressing the conditioned $Y$ in this form means that we can still express our conditioned quantum process following the derivation in Ref.~\cite{Albash2025} as:
\begin{eqnarray} \label{eqt:CumulantK}
\mathcal{K}(t,t_0) &=& -i \ev{ \tilde{\mathcal{L}}_1(t,t_0) } - i \ev{  \tilde{\mathcal{L}}_2(t,t_0) } \nonumber \\
&& - \frac{1}{2} \left( \ev{\tilde{\mathcal{L}_1}^2} -  \ev{\tilde{\mathcal{L}_1}}^2 \right) + \dots \nonumber \\
&=& -i { \tilde{\mathcal{L}}_1^{(D)}(t,t_0) } - i {  \tilde{\mathcal{L}}_2^{(D)}(t,t_0) } \nonumber \\
&& \hspace{-1.5cm} - i \ev{  \tilde{\mathcal{L}}_2^{(S)}(t,t_0) } - \frac{1}{2} \left( \ev{ \left( \tilde{\mathcal{L}}_1^{(S)} \right)^2} \right) + \dots \ , 
\end{eqnarray}
where the superscript $S$ terms only depend on the expectation values of the primed stochastic process $Y'$.
  
We can now identify the contributions to the various terms in the above equation (Eq.~\eqref{eqt:CumulantK}). 
For convenience, let us define the functions:
\begin{eqnarray}
f_{n}(t) &=&   \mathbb{E} \left[ \left(X_n(t) | \partial_n  \right) \right] = x_{k-1} e^{-\gamma_n (t-t_{k-1})} \nonumber \\
&& + \frac{e^{2 \gamma_n(t-t_{k-1})}-1}{e^{2 \gamma_n \Delta t}-1} \left( e^{\gamma_n (t_k- t)} x_k \right. \nonumber \\
&& \left. -  e^{-\gamma_n(t-t_{k-1})} x_{k-1} \right) \ , \\
g_n(s,t) &=&  \mathrm{E} \left[  X'_n(t) X'_n(s) \right]  \nonumber \\
&=&  \frac{\sigma_n^2}{\gamma_n} \frac{ \sinh(\gamma_n (s - t_{k-1})) \sinh( \gamma_n (t_k-t))}{\sinh(\gamma_n \Delta t)} \ , \nonumber \\
\end{eqnarray}
with $\mathrm{Var} \left[ X'_n(t)  \right] = g_n(t,t)$. We note that $f_n$ is first order in the strength of the OU noise process, while $g_n$ is second order in the strength of the OU noise process. The contribution to $-i { \tilde{\mathcal{L}}_1^{(D)}(t,t_0) }$ comes from:
\begin{eqnarray}
\mathbb{E} \left[ \eta | \partial \right] &=& \mathbb{E} \left[ Y(t) | \partial  \right] - 1 \nonumber \\
&\approx&  \sum_{n} \left( f_{n}(t) + \frac{1}{2} g_n(t,t)  \right) + \frac{1}{2} \sum_{m, n} f_m(t) f_n(t) \ . \nonumber \\
\end{eqnarray}
The first term corresponds to using the approximation $\eta \approx \delta V / \mathcal{I}$, but we see that by not making this approximation we get two additional second order contributions in the strength of the OU noise process at first order in the Magnus expansion.

Because the remaining terms, $- i {  \tilde{\mathcal{L}}_2^{(D)}(t,t_0) }, - i \ev{  \tilde{\mathcal{L}}_2^{(S)}(t,t_0) } ,- \frac{1}{2} \ev{ \left( \tilde{\mathcal{L}}_1^{(S)} \right)^2} $ are quadratic in the stochastic noise processes and if we keep only terms that are second order in the strength of the OU noise processes, we get the same contribution as using the approximation $\eta \approx \delta V / \mathcal{I}$.

For our noise Hamiltonian, each $\delta V_{ij}(t)$ associated with each exchange axis is taken to be a sum of 8 OU processes with frequencies distributed log-linearly~\cite{Kaulakys2005} between 1 mHz and 1 GHz. Each OU process has a PSD of the form in Eq.~\eqref{eqt:OUpsd}. We fix the insensitivity to $\mathcal{I} = 10$mV. In order to achieve a free induction decay with a desired $T_2^\ast$ value with an exchange of 100MHz~\cite{Andrews2019}, we choose $p$ values according to Table~\ref{table:voltageT2}. Further details about OU processes relevant to this construction can be found in Ref.~\cite{Albash2025}.
\begin{table}[!htbp]
\centering
\begin{tblr}{
  colspec = {|c|c|c|},
}
\hline
  Noise model label & $T_2^\ast$ ($\mu$s)     &    $p$  (mV$^2$) \\
  \hline[2pt]
\cline{1-4}
NM1 & $0.5$ & $6.92857 \times 10^{-4}$ \\
NM2 & $1.0$ & $1.78442 \times 10^{-4}$\\
NM3 & $2.0$ & $4.59540 \times 10^{-5}$\\
\hline
\end{tblr}
\caption{Voltage noise OU parameter $p$ to achieve a desired free induction decay $T_2^\ast$ value at an exchange interaction strength of 100MHz and a 10mV insensitivity.} \label{table:voltageT2}
\end{table}
\begin{figure}[htbp] 
   \centering
   \includegraphics[width=3.25in]{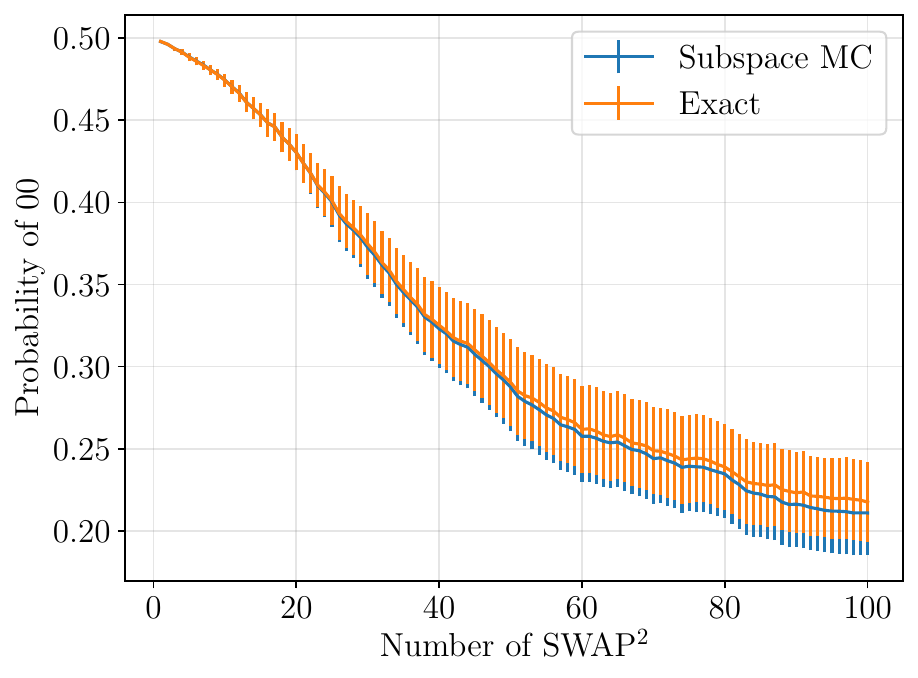}  
   \caption{\textbf{Repeated SWAP pairs with Pauli randomized compiling.} Probability of measuring the outcome 0 for both qubits  when simulating the circuit in Fig.~\ref{fig:RepeatedRCSwapsCircuit} but with Pauli gates. The initial state is the Bell state $\ket{\Phi} = \frac{1}{\sqrt{2}} \left( \ket{00} + \ket{11} \right)$. Simulation results are averaged over 100 noise realizations using the noise model NM1, with the error bars being the $2\sigma$ confidence intervals generated by performing a bootstrap over the 100 noise realizations. For the Subspace MC simulations, 50 MC trials are performed per noise realization.}\label{fig:SwapPauli}
\end{figure}
\begin{figure}[htbp] 
   \centering
   \includegraphics[width=3.25in]{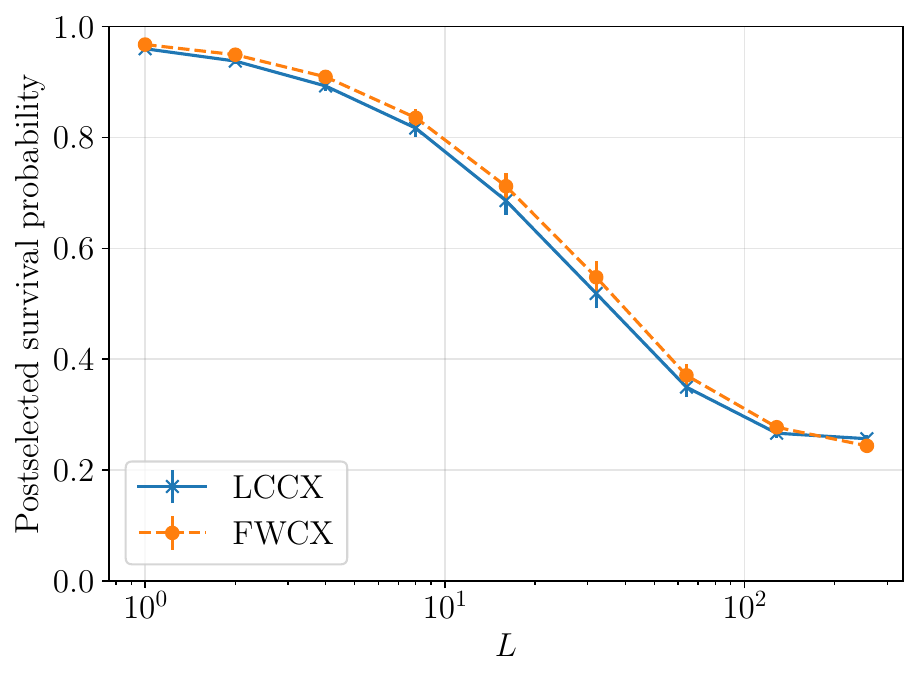}  
   \caption{\textbf{Two-qubit RB with leakage post-selection.} The post-selected survival probability as a function of circuit depth $L$ for 2-qubit randomized benchmarking with RiL gadgets as depicted in Fig.~\ref{fig:2qRBwithRiL}. The results are post-selected on the two RiL gadgets detecting no leakage events on the input qubits. At every depth, 50 random circuits are averaged over. The results are averaged over 40 independent noise realizations {using noise model NM1}, and 5 Monte Carlo trials are performed per noise realizations. Error bars are $2\sigma$ confidence intervals generated by performing a bootstrap over the 40 noise realizations. }\label{fig:2QRBwithRiLsurvival}
\end{figure}
\section{Additional Results}  \label{app:MoreResults}
%
In Fig.~\ref{fig:SwapPauli}, we show the results for repeated SWAP operations as discussed in Sec.~\ref{sec:Swap} but now using single-qubit Pauli gates for the randomized compiling. We again find agreement between the exact and Subspace MC simulations within statistical confidence.

To support the results in Fig.~\ref{fig:2QRBwithRiL}, we show in Fig.~\ref{fig:2QRBwithRiLsurvival} the survival probability for 2-qubit randomized benchmarking post-selected on measurement outcomes with no leakage. The survival probability is very similar for circuits using either type of CNOT gate because we chose to follow each 2-qubit Clifford operation with an idle of 400ns.

 \section{RiL}
 In Ref.~\cite{Langrock2020}, a procedure to identify reset-if-leaked (RiL) gates for exchange-only qubits is presented. Using a single ancilla qubit (only 2 spins of the 3 spins are required) prepared in the singlet state, the flagged-RiL gate operates ideally on a data qubit by:
 (a) applying the identity operation if the qubit state is in the computational subspace, (b) reseting the qubit state to a fixed state if the qubit state is in the leakage subspace, with a measurement of the ancilla qubit flagging which has occurred. If the ancilla is measured to be a singlet state (outcome 0), then case (a) has occurred, whereas if the ancilla is measured to be in a triplet state (outcome 1), then case (b) has occurred. We provide different flagged-RiL gates that reset to different states below, where we have assumed that the gate operates as:
 \beq
\hat{U} = \left( e^{-i \Theta_{12} \vec{\hat{S}}_1 \cdot \vec{\hat{S}}_2} e^{-i \Theta_{34} \vec{\hat{S}}_3 \cdot \vec{\hat{S}}_4} e^{-i \Theta_{01} \vec{\hat{S}}_0 \cdot \vec{\hat{S}}_1} e^{-i \Theta_{23} \vec{\hat{S}}_2 \cdot \vec{\hat{S}}_3} \right)^m \ .
 \eeq

 \begin{table}[!htbp]
\centering
\begin{tblr}{
  colspec = {c|c|c|c},
  cell{1}{1} = {c=4}{c},
}
Reset to $0.7873 \ket{0} + 0.6166 e^{-0.5113 \pi i}\ket{1} $~\cite{Langrock2020}\\
\hline[2pt]
  $\Theta_{23}/\pi$     &   $\Theta_{01}/\pi$ &  $\Theta_{34}/\pi$ & $\Theta_{12}/\pi$  \\
\cline{1-4}
0.496474 & 0.511053 & 0.407919 & 1.128462 \\
0.644573 & 1.456051 & 0.233065 & 1.473077 \\
1.574455 & 1.481738 & 0.296057 & 0.778243 \\
0.458866 & 0.762262 & 0.654983 & 0.907327 \\
0.495382 & 0.403991 & 0. & 1.700957 \\
\hline[2pt]
\end{tblr}
\end{table}
 \begin{table}[!htbp]
\centering
\begin{tblr}{
  colspec = {c|c|c|c},
  cell{1}{1} = {c=4}{c},
}
Reset to $\ket{0}$ \\
\hline[2pt]
  $\Theta_{23}/\pi$     &   $\Theta_{01}/\pi$ &  $\Theta_{34}/\pi$ & $\Theta_{12}/\pi$  \\
\cline{1-4}
0.404872 & 1.26179 & 0.662913 & 1.322802 \\
0.848643 & 0.61924 & 0.007398 & 1.314729 \\
1.490594 & 1.695266 & 0.511201 & 0.93564 \\
0.436651 & 1.062405 & 0.764366 & 0.239862 \\
0.146759 & 0.019509 & 0.002173 & 1.473893 \\
1.249776 & 0.295492 & 0.000241 & 0.010045 \\
\hline[2pt]
\end{tblr}
\end{table}
  \begin{table}[!htbp]
\centering
\begin{tblr}{
  colspec = {c|c|c|c},
  cell{1}{1} = {c=4}{c},
}

Reset to $\frac{1}{\sqrt{2}} \left( \ket{0} -i \ket{1} \right) $ \\ 
\hline[2pt]
  $\Theta_{23}/\pi$     &   $\Theta_{01}/\pi$ &  $\Theta_{34}/\pi$ & $\Theta_{12}/\pi$  \\
\cline{1-4}
0.886238 & 1.709463 & 1.015384 & 1.710426 \\
0.119201 & 1.371027 & 1.423231 & 0.504354 \\
1.467197 & 0.92291 & 0.643456 & 0.927318 \\
1.013166 & 0.2013 & 1.667896 & 0.785257 \\
1.463585 & 0.469296 & 0.371406 & 1.528549 \\
1.279473 & 1.202639 & 0.577226 & 1.449496 \\
0.663838 & 6.726766e-05 & 1.280471 & 1.310563 \\
\hline[2pt]
\end{tblr}
\end{table}
  \begin{table}[!htbp]
\centering
\begin{tblr}{
  colspec = {c|c|c|c},
  cell{1}{1} = {c=4}{c},
}
Reset to $\frac{1}{\sqrt{2}} \left( \ket{0}  + i \ket{1} \right) $ \\ 
\hline[2pt]
  $\Theta_{23}/\pi$     &   $\Theta_{01}/\pi$ &  $\Theta_{34}/\pi$ & $\Theta_{12}/\pi$  \\
\cline{1-4}
0.367425 & 0.390655 & 0.675156 & 1.715356 \\
0.088007 & 1.378581 & 0.059881 & 0.60918 \\
1.597326 & 1.556397 & 0.359343 & 1.047532 \\
0.965406 & 0.230674 & 1.554678 & 1.724771 \\
1.165393 & 0.330036 & 0.614287 & 1.365049 \\
1.747609 & 0.614656 & 0.58181 & 1.66819 \\
0.475027 & 6.379607e-07 & 1.770441 & 1.645198 \\
\hline[2pt]
\end{tblr}
\end{table}
  \begin{table}[!htbp]
\centering
\begin{tblr}{
  colspec = {c|c|c|c},
  cell{1}{1} = {c=4}{c},
}
Reset to $ \ket{1} $ \\ 
\hline[2pt]
  $\Theta_{23}/\pi$     &   $\Theta_{01}/\pi$ &  $\Theta_{34}/\pi$ & $\Theta_{12}/\pi$  \\
\cline{1-4}
0.751282 & 1.339536 & 0.544124 & 1.800207 \\
0.475031 & 1.018374 & 1.678032 & 0.687548 \\
1.146998 & 1.330291 & 0.474679 & 0.845616 \\
0.588767 & 0.481059 & 0.012724 & 1.412538 \\
1.829759 & 0.27343 & 0.002325 & 1.120526 \\
1.059034 & 1.579138 & 0.565933 & 1.477051 \\
0.450102 & 1.999846 & 0.152572 & 1.490437 \\
\hline[2pt]
\end{tblr}
\end{table}
  \begin{table}[!htbp]
\centering
\begin{tblr}{
  colspec = {c|c|c|c},
  cell{1}{1} = {c=4}{c},
}
Reset to $\frac{1}{\sqrt{2}} \left( \ket{0} + \ket{1} \right) $ \\ 
\hline[2pt]
  $\Theta_{23}/\pi$     &   $\Theta_{01}/\pi$ &  $\Theta_{34}/\pi$ & $\Theta_{12}/\pi$  \\
\cline{1-4}
0.374601 & 0.860477 & 0.998716 & 1.36291 \\
0.480212 & 1.369092 & 1.598713 & 0.568627 \\
1.295135 & 1.382194 & 0.696801 & 0.556663 \\
1.166426 & 0.909913 & 0.960929 & 1.518589 \\
0.099706 & 0.441881 & 0.275423 & 1.699717 \\
1.151868 & 0.977891 & 0.602636 & 1.456722 \\
0.667264 & 1.742269 & 1.17429 & 1.434421 \\
\hline[2pt]
\end{tblr}
\end{table}
  \begin{table}[!htbp]
\centering
\begin{tblr}{
  colspec = {c|c|c|c},
  cell{1}{1} = {c=4}{c},
}
Reset to $\frac{1}{\sqrt{2}} \left( \ket{0} - \ket{1} \right) $ \\ 
\hline[2pt]
  $\Theta_{23}/\pi$     &   $\Theta_{01}/\pi$ &  $\Theta_{34}/\pi$ & $\Theta_{12}/\pi$  \\
\cline{1-4}
0.884353 & 1.982832 & 0.77444 & 1.568513 \\
0.325984 & 0.95356 & 1.572626 & 0.363891 \\
1.253394 & 1.196483 & 0.411544 & 1.11218 \\
0.900993 & 0.705653 & 1.569746 & 0.941712 \\
1.594288 & 0.482192 & 0.124673 & 1.596073 \\
0.870418 & 1.346642 & 0.507325 & 1.162075 \\
0.539638 & 1.690092 & 0.219089 & 1.50859 \\
\hline[2pt]
\end{tblr}
\end{table}

As mentioned in the main text, the ideal operation of the RiL gate can generate superpositions of different states with different spin projection quantum numbers for the qubits even when the input qubits each have a definite spin projection quantum number, in contrast to logical operations. To illustrate this, we take the reset-to-$\ket{0}$ RiL gate, and we assume the input state is given by $ \ket{\frac{3}{2}, 1,  \frac{1}{2}} \ket{\frac{1}{2}, 0, \frac{1}{2}}$ (NZ-ZN orientation). This input state has a leaked state on the data qubit, with $S_z$ value $-1/2$ and the ancilla qubit is set to $\ket{0}$ with $S_z$ value $1/2$. For the reset-to-$\ket{0}$ RiL gate, the output state is approximately given by:
\begin{eqnarray}
\hat{U}_{\mathrm{RiL}} \ket{\frac{3}{2}, 1,  \frac{1}{2}} \ket{\frac{1}{2}, 0, \frac{1}{2}} &=& \nonumber \\
&& \hspace{-3.5cm} \left( -0.3853 -0.2716i \right) \ket{\frac{1}{2}, 0, \frac{1}{2}} \ket{\frac{1}{2}, 1, \frac{1}{2}} \nonumber \\
&& \hspace{-3.5cm} + \left( 0.5449 + 0.3842i \right)  \ket{\frac{1}{2}, 0, \frac{1}{2}}  \ket{\frac{3}{2}, 1, \frac{1}{2}} \nonumber \\
&& \hspace{-3.5cm} + \left( 0.4719 + 0.3327i \right)  \ket{\frac{1}{2}, 0, -\frac{1}{2}}  \ket{\frac{3}{2}, 1, \frac{3}{2}} \ ,
\end{eqnarray}
where the data qubit state now includes a superposition of $1/2$ and $-1/2$ {spin projection quantum number} states. 
%
\section{Layout for parity check circuit}  \label{app:Layout}
\begin{figure}[h] %
   \centering
   \includegraphics[width=1.8in]{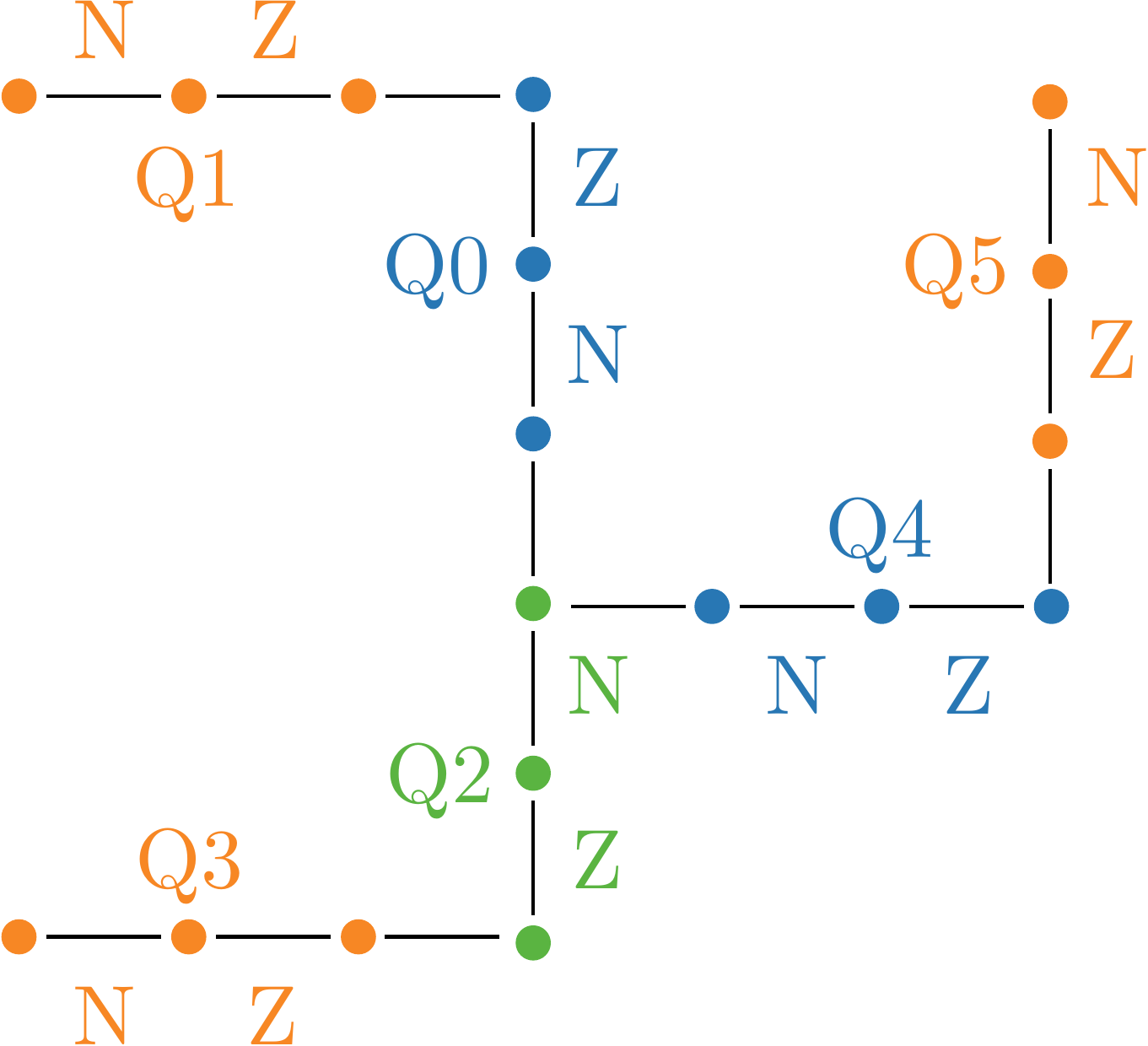} 
   \caption{\textbf{Illustration of the qubit layout for the parity check circuits.} Each EO qubit covers three {vertices}, corresponding to an electron/dot, and the solid lines connecting them correspond to exchange axes. The EO qubits are labeled Q0 to Q5 with the numbering corresponding to the qubits in Fig.~\ref{fig:ParityCheckwithRiL} from top to bottom, and the $-z$ rotation and $n$ rotation axes for each EO qubit are labeled as Z and N respectively. The RiL ancilla qubits (Q1, Q3, Q5) are shown in orange, the data qubits (Q0, Q4) are shown in blue, and the measurement ancilla qubit (Q2) is shown in green.}
   \label{fig:DotLayout}
\end{figure}

%
In order to minimize the number of SWAP operations we need to perform in our parity check circuits, we choose a layout and orientations for the qubits as illustrated in Fig.~\ref{fig:DotLayout}.
\clearpage
\end{document}